%% file: main.tex
\documentclass{vldb}
\usepackage{graphicx}

\usepackage{amsthm}

\usepackage[noend]{algorithmic}
\usepackage{algorithm}
\usepackage{balance}  % for  \balance command ON LAST PAGE  (only there!)

\usepackage{comment}

\newtheorem{definition}{Definition}
\newtheorem{property}{Property}
\newtheorem{theorem}{Theorem}
\newtheorem{lemma}{Lemma}

\vldbTitle{Dynamic Skyline Queries on Encrypted Data Using Result Materialization}
\vldbAuthors{Sepanta Zeighami, Gabriel Ghinita, Cyrus Shahabi}
\vldbDOI{https://doi.org/10.14778/xxxxxxx.xxxxxxx}
\vldbVolume{XX}
\vldbNumber{xxx}
\vldbYear{2020}

\begin{document}

% ****************** TITLE ****************************************

\title{Secure Dynamic Skyline Queries \\ Using Result Materialization}

\numberofauthors{3}

\author{
\alignauthor
Sepanta Zeighami \\
       \affaddr{University of Southern California}\\
       \email{zeighami@usc.edu}
\alignauthor
Gabriel Ghinita\\
       \affaddr{University of Massachusetts Boston}\\
       \email{gabriel.ghinita@umb.edu}
\alignauthor Cyrus Shahabi\\
       \affaddr{University of Southern California}\\
       \email{shahabi@usc.edu}
}

\maketitle 

\begin{abstract}
%skyline is increasingly popular and important; it id deployed in the cloud and existing approaches have limitations
Skyline computation is an increasingly popular query, with broad applicability in domains such as healthcare, travel and finance. Given the recent trend to outsource databases and query evaluation, and due to the proprietary and sometimes highly sensitivity nature of the data (e.g., in healthcare), it is essential to evaluate skylines on encrypted datasets. Several research efforts acknowledged the importance of secure skyline computation, but existing solutions suffer from at least one of the following shortcomings: (i) they only provide ad-hoc security; (ii) they are prohibitively expensive; or (iii) they rely on unrealistic assumptions, such as the presence of multiple non-colluding parties in the protocol. 

%following similar work in NN, we do pre-computation; we provide thorough theoretical treatment and show that it performs well
Inspired from solutions for secure nearest-neighbors (NN) computation, we conjecture that the most secure and efficient way to compute skylines is through result materialization. However, this approach is significantly more challenging for skylines than for NN queries. We exhaustively study and provide algorithms for pre-computation of skyline results, and we perform an in-depth theoretical analysis of this process. We show that pre-computing results while minimizing storage overhead is NP-hard, and we provide dynamic programming and greedy heuristics that solve the problem more efficiently, while maintaining storage at reasonable levels. Our algorithms are novel and applicable to plain-text skyline computation, but we focus on the encrypted setting where materialization reduces the cost of skyline computation from hours to seconds. Extensive experiments show that we clearly outperform existing work in terms of performance, and our security analysis proves that we obtain a smaller (and quantifiable) data leakage than competitors.
\end{abstract}

\section{Introduction}\label{sec:intro}
\input{Introduction.tex}

\section{Background and Definitions}\label{sec:setting}
\input{Definitions.tex}

\input{Tiling.tex}

\input{TileAggregation.tex}

\input{GenTiling.tex}

\input{FinalAlgo.tex}

\input{Encryption.tex}

\section{Empirical Evaluation}\label{sec:exp}
\input{Exp.tex}

\section{Related Work}\label{sec:related}
\input{Related.tex}

\section{Conclusion}\label{sec:conclusion}
\input{Conclusion.tex}

\bibliographystyle{abbrv}
\bibliography{references}  % vldb_sample.bib is the name of the Bibliography in this case

\clearpage
\appendix
\input{Appx.tex}

\end{document}

%% file: Introduction.tex
%Consider a point $p\in D$. For a query $q$, $p$ is in the skyline if $p$ is not dominated by any other point in $D$ with respect to $q$. However, $q$ is not known in advance. Existing methods spend a large amount of computation on determining whether $p$ is dominated by any other point after obtaining $q$ from the user. However, such a computation becomes prohibitive if the data is encrypted. 

%%===
The skyline query finds points in a dataset which are not dominated by any other data point in at least one attribute value. These points have the property of ``standing out'' among other data points. For instance, in airfare booking, the skyline may contain routes that are either cheapest, shortest, or have the fewest stopovers. In a hospital database, the skyline may contain patients with lowest age, or patients with minimum value for a certain test result (e.g., hemoglobin level). Many research efforts in the past decade focused on efficient computation of skylines over plaintext data~\cite{borzsony2001skyline,Kossman02,chan2006finding}. However, few solutions exist for the problem of {\em secure} skyline, where the data and query execution are outsourced to a service provider (SP). Since the data may be proprietary or protected by law (e.g., healthcare records), the computation must be executed over {\em encrypted} datasets. 

The work in~\cite{Bothe14} was the first to formulate the secure skyline problem, but the solution proposed only provided ad-hoc security. Later in \cite{FGCS16,liu2018secure,IoT19}, several solutions were proposed that used either homomorphic encryption, or secure multi-party computation. However, they either leak excessive information to the SP, or they incur prohibitive computation and communication costs. The state-of-the-art approach in~\cite{liu2018secure} assumes a system architecture with two non-colluding parties that engage in a secure multi-party protocol that needs to scan the entire dataset for each query, and perform expensive operations for a large subset of the Cartesian product of all records. This results in a response time of around 3 hours for a single query, which is clearly impractical. Each query starts anew, and cannot use any information computed form the previous query.

We propose a different approach, which has been shown in the seminal work of \cite{SNN} to be the only secure and efficient approach for computing nearest-neighbor (NN) queries on encrypted data. The main idea of \cite{SNN} is to pre-compute query results using a Voronoi diagram, and then partition and materialize the results in a data structure whose properties are not dependent on the data characteristics (to minimize leakage). At query time, the user provides an encrypted representation of the query point, and then the SP and the user engage in an interactive protocol that allows the user to retrieve the partition that contains the results to the query (together with a number of possible additional results, i.e., false positives)\footnote{The Voronoi diagram materialization used in SNN was first introduced by the authors of this submission in \cite{GKKST08} for private NN queries on public datasets.}. It is shown in~\cite{SNN} that any method that uses some sort of encrypted processing directly on the data points leaks a significant amount of information, in the form of either inter-point distances, or distance order.

Inspired by~\cite{SNN}, we extend the idea of pre-computation and partitioning to the skyline query. This leads to a more secure solution, and a much faster response time, as we do {\em not} process the entire dataset for each query. However, pre-computing skyline results in the dynamic case (i.e., for every possible query point) is a very challenging problem. In fact, the equivalent of this problem on {\em plaintext} has attracted limited attention, because doing so would be too expensive. Contrast this to the case of NN queries, where Voronoi diagrams have been extensively used even for plaintext queries. Our work is pioneering in that it provides a thorough analysis of skyline result pre-computation, which may find applications beyond encrypted data. Although expensive in comparison to other plaintext skyline computation counterparts, we believe that the pre-computation approach becomes valuable, and in fact the only viable approach, in the context of encrypted data. This is because techniques that do not perform materialization require hours of processing for a single query. Our approach can answer a query in less than a second, even though there is a one-time setup cost.

In a nutshell, our materialization approach reduces the skyline query to a simple index look-up. First, we perform a partitioning of the space into non-overlapping regions called {\em skyline tiles}, such that the answer to all skyline queries that fall within the same tile are identical. This is done by finding, for each data point, the space of queries for which the data point is in the skyline (see Figs.~\ref{fig:DomRegions} and \ref{fig:SkyRegions}). Then, skyline tiles are created by intersecting these regions (see Fig.~\ref{sec:tiling}). We store the query answer in each tile, and we index the tiles in a data structure. This way, we can answer a dynamic skyline query by simply performing a lookup in the index. The index is then encrypted to ensure the security of our approach. Finally, note that reducing skyline query computation to an index look-up allows us to utilize significantly less expensive cryptographic primitives, which manifests itself in orders of magnitude improvement in query time compared with existing approaches.

%Inspired from solutions for secure nearest-neighbors (NN) computation, we conjecture that the most secure and efficient way to compute skylines is through result materialization and data domain partitioning. However, taking this approach is significantly more challenging for skylines than for NN queries. . . Our experimental evaluation prove that , and our security analysis shows that we obtain a smaller (and quantifiable) data leakage than competitors.

Our specific contributions are:
\begin{enumerate}
\item
We comprehensively study the problem of pre-computing skyline results, and we perform an in-depth theoretical analysis of this process.

\item
We show that the problem of pre-computing results while minimizing storage overhead is NP-hard, and we provide dynamic programming and greedy heuristics that solve the problem more efficiently, while maintaining storage costs at reasonable levels.

\item
We perform an extensive experimental evaluation showing that our techniques clearly outperform existing work in terms of performance.

\item
We provide an in-depth security analysis to measure the leakage of our proposed approach, and conclude that the amount of leakage is quantifiable, and smaller compared to existing techniques.
\end{enumerate}

The rest of the paper is organized as follows: we provide background information in Section 2. Section 3 introduces a construction to pre-compute skyline query results for data units called {\em tiles}. We show how to aggregate tiles and perform data partitioning to reduce storage overhead in Section 4. We generalize the tile concept in Section 5 to obtain further performance gains, and outline the complete, end-to-end solution to the secure skyline query in Section 6. Section 7 presents our security analysis, followed by experiments in Section 8. We review related work in Section 9 and conclude in Section 10.

%% file: Definitions.tex
\subsection{Preliminaries}

Consider database $D$ with $n$ points in the $d$-dimensional space $R^d$. For point $p\in D$, $p[i]$ denotes its value in $i^{th}$ dimension. A query is denoted by $q\in R^d$. For ease of discussion, let $\infty$ be a large constant and assume that the query space is bounded by $\infty$ in every dimension (i.e., $q[i]<\infty$ for all $i$). The {\em domination} relationship between two points in $D$ with respect to a query $q$ is defined as follows:
\begin{definition}[Domination]
A point $p\in D$ dominates another point $p'\in D$ with respect to $q$, denoted by $p>_q p'$, if and only if $\forall i, 1\leq i\leq d, |p[i]-q[i]| \leq |p'[i]-q[i]|$ and $\exists i, 1\leq i\leq d, |p[i]-q[i]|< |p'[i]-q[i]|$. We use $p\not>_q p'$ if $p$ does not dominate $p'$ with respect to $q$.
\end{definition}

Intuitively, $p>_q p'$ implies that $p$ is at least as close to $q$ as $p'$ in all dimensions, and it is closer to $q$ in at least one. 
%This domination relationship reduces to the traditional skyline domination relationship if $q$ is the origin. 
%Next, we define the dynamic skyline query.

\begin{definition}[Dynamic Skyline Query]
Given a database $D$ and a query point $q$, return a set $S\subseteq D$, such that no point in $S$ is dominated by a point in $D$ with respect to $q$, that is, $S = \{p\in D| \forall p'\in D, p' \not>_q p\}$.
\end{definition}

\begin{figure*}[!ht]
    \centering
    \includegraphics[width=\textwidth]{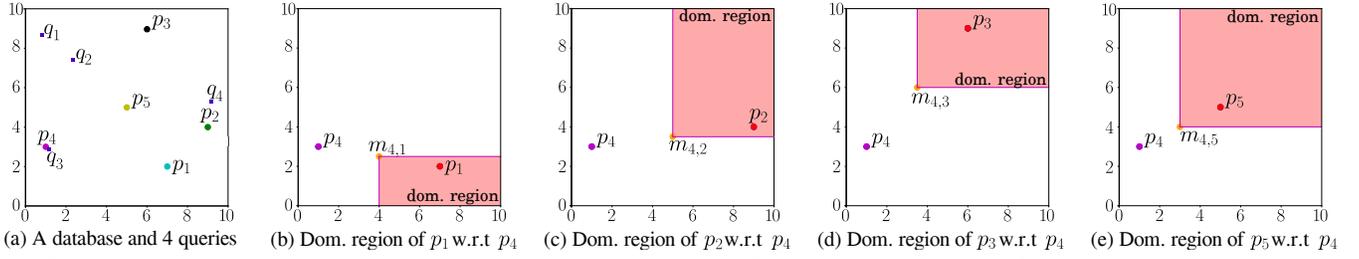}
    \vspace{-20pt}
    \caption{(a) A database and 4 sample queries. (b) - (e) Domination regions of all the points w.r.t $p_4$}
    \vspace{-10pt}
    \label{fig:DomRegions}
\end{figure*}

Note that, the conventional skyline query (where $q$ is the domain origin) can be defined as a special case of dynamic skyline. Our focus is on the more challenging dynamic skyline setting, so whenever we mention skyline query, we refer to the dynamic case (given query $q$). %, as opposed to the conventional skyline where $q$ is the origin. 
We use Fig. \ref{fig:DomRegions} (a) as a running example: the skyline query answers for queries $q_1$ and $q_2$ are $\{ p_3, p_4, p_5\}$, and $\{p_2,p_5\}$, respectively.

\subsection{System Model}
We assume three types of participants: the data owner (DO), the service provider (SP), and users. Users are trusted by the DO, and wish to obtain the result to dynamic skyline queries on the dataset owned by DO. The DO does not have the infrastructure to run such a service, so it outsources the functionality to SP (e.g., a commercial entity), which is {\em honest but curious}. The SP runs the protocol correctly, but may try to infer private details about the data. In addition, the SP may be compromised by an attacker, in which case the data may be exposed, with serious consequences (e.g., leakage of healthcare records). To address such threats, the DO first encrypts the dataset, and only shares the encrypted version with the SP. At runtime, the DO may be offline, and only the SP and the user engage in a protocol to determine the encrypted result to the user's skyline query $q$. 

Users are trusted with some secret tokens (e.g., encryption keys), and are assumed not to collude with the SP (in practice, the users may be highly vetted individuals, e.g., medical doctors). The user retrieves a superset of the actual query result in encrypted form (i.e., including false positives), and performs a local {\em lightweight} filtering step to narrow down the exact result. Our solution {\em guarantees} that the user always obtains the exact result, and we also provide an upper bound on the total amount of false positives, in order to minimize the communication and computational cost on the user.

To support this protocol, the DO must prepare and encrypt the dataset, which may incur a significant overhead. However, this is a {\em one-time setup cost}. It helps reduce significantly the query processing overhead at runtime, which is the most important component of the cost, since that is the response time perceived by the user. We also assume that when the user registers for the service, there is a one-time setup cost on the user device. This may include transferring of a relatively small amount of metadata needed to run the protocol with the SP (our evaluation shows that the user download size is in the order of $10s$ of MB, which is a reasonable amount even on a mobile connection). 

%We aim at answer the dynamic skyline query in the absence of a trusted server. We assume the data owner encrypts the data and the meta data, encrypted queries are sent to the server who is not trusted by the clients or the data owners, and the server uses the encrypted data and the encrypted query to return the dynamic skyline of the database with respect to the query point.  

\begin{figure*}[!ht]
    \centering
    \includegraphics[width=\textwidth]{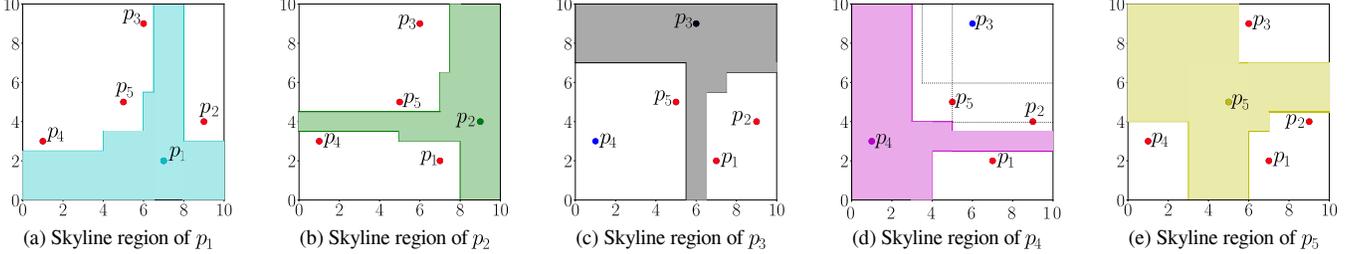}
    \vspace{-20pt}
    \caption{Skyline regions of all points}
    \label{fig:SkyRegions}
    \vspace{-10pt}
\end{figure*}

%\subsection{Background on Encryption}
%[TODO Gabriel. Anticipated at 3/4 of a column, or perhaps combine with Section 7 to save space]

%% file: Tiling.tex
\begin{table}[t]
    \centering
    \begin{tabular}{c|c}
        {\bf Symbol} & {\bf Definition} \\\hline
        $D$, $n$, $d$ & $d$-dimensional database $D$ of cardinality $n$ \\\hline
        $p>_q p'$ & $p$ dominates $p'$ with respect to query $q$\\\hline
        $D_{p'}^{p}$ & Domination region of $p'$ with respect to $p$ \\\hline
        $S_{p}$ & Skyline region of $p$ \\\hline
        $m_{p, p'}$ & Mid-point between $p$ and $p'$ \\\hline
        $cnt(T)$, $spc(T)$ & For $T=(S,P)$, $cnt(T)=P$, $spc(T)=S$ \\\hline
        $\mathcal{T}_{D}$ & Set of skyline tiles for database $D$ \\\hline
        $L_i$ & Boundaries of skyline tiles in dim. $i$ \\\hline
        $N_i$, $N$ & $N_i=|L_i|$, $N=\max_i\{N_i\}$ \\\hline
        $\mathcal{I}$ & Set of skyline indices \\\hline
        $l$ & Number of skyline regions to intersect \\\hline
        $k$ & Max. number of false positives allowed \\\hline
        $m$ & Pre-partitioning parameter \\\hline
    \end{tabular}
    \vspace{-10pt}
    \caption{Summary of Notations}
    \vspace{-15pt}
    \label{tab:notation}
\end{table}

\section{Skyline Result Materialization}\label{sec:tiling}
In this section, we introduce some preliminary concepts that are built upon later in Sections~\ref{sec:agg} and~\ref{sec:genTiles} to obtain efficient algorithms for building and storing an index on the plaintext data. Section~\ref{sec:algo} presents a complete, end-to-end processing algorithm on plaintext data. Finally, in Section~\ref{sec:encrypteion} we show how the index is encrypted using a special transformation before being sent to the SP, and how traversal is performed on the encrypted structure.

%We first introduce the \textit{skyline region} of a point and then describe how it can be used to partition the space and answer the skyline query. 

\subsection{Skyline Region of a Point}
Consider a point $p\in D$. Recall that for a query $q$, $p$ is in the skyline if $p$ is not dominated by any other point in $D$ with respect to $q$. Denoted by $S_p$ is the \textit{skyline region} of $p$, i.e., the set of all query points for which $p$ is a skyline point: $S_p = \{x\in R^d| \forall p'\in D, p' \not>_x p\}$. Due to the properties of the domination relation, $S_p$ is a polytope of a specific shape and can be constructed easily. We introduce several auxiliary concepts needed to define skyline regions.

\textbf{Domination Region of a point}. First, consider two points $p$ and $p'$. Recall that $p'>_q p$, if and only if 
\begin{align}\label{ineq:forall}
\forall_{i, 1\leq i\leq d}\; |p'[i]-q[i]| \leq |p[i]-q[i]|
\end{align} 
and 
\begin{align}\label{ineq:exists}
\exists_{i, 1\leq i\leq d}\; |p[i]-q[i]|< |p[i]-q[i]|
\end{align} 
Rephrasing the definition of domination, observe that $p'$ dominates $p$ for all the query points $q$ in $D_{p'}^{p} = \{q\in R^d| q\; $ satisfies (\ref{ineq:forall}) and (\ref{ineq:exists})$\}$. 
%In other words, $p'$ dominates $p$ with respect to all query points $q$ such that $p'$ is at least as close as $p$ to $q$ in all the dimensions and closer to $q$ in at least one dimension. 
We refer to $D_{p'}^p$ as the domination region of $p'$ with respect to $p$. For convenience, define $D_{p}^p=\emptyset$. The dominance region of all the points with respect to $p_4$ is shown in Fig.~\ref{fig:DomRegions} (b)-(e). 

Note that $D_{p'}^p$ is the solution to inequalities (\ref{ineq:forall}) and (\ref{ineq:exists}). We first focus on the solutions to Eq.~\eqref{ineq:forall}. Observe that 
\begin{align}\label{ineq:domregion}
|p'[i]-q[i]| \leq |p[i]-q[i]|\Longleftrightarrow\left\{\begin{matrix}
 q[i] \geq \frac{p'[i]+p[i]}{2} & p'[i]>p[i] \\ 
 q[i] \leq \frac{p'[i]+p[i]}{2} & p'[i]<p[i] \\ 
 \text{true} & \text{otherwise} 
\end{matrix}\right.
\end{align}
Given $p$ and $p'$ and for each $i$, (\ref{ineq:domregion}) is an inequality of the form $q[i]\leq c$ or $q[i]\geq c$, for some constant $c$ depending on $p[i]$ and $p'[i]$. That is, the solution to the inequality for each $i$ is of the form $(-\infty, c]$ or $[c, \infty)$. Let $m_{p',p}$ be the mid-point between $p$ and $p'$, i.e., $m_{p', p}[i] = \frac{p'[i]+p[i]}{2}$. Based on Eq.~\eqref{ineq:domregion}, a $q\in R^d$ satisfies Eq.~\eqref{ineq:forall} if
\vspace{-5pt}
\begin{align}\label{eq:domregion2}
\forall_{i, 1\leq i\leq d}, q[i] \in \left\{\begin{matrix}
 [m_{p, p'}[i], \infty)  & p'[i]>p[i] \\ 
 (-\infty, m_{p, p'}[i]] & p'[i]<p[i] \\
 (-\infty, \infty)  & p'[i]=p[i] 
 \end{matrix}\right.
\end{align}

Let $Z_{p'}^p$ be the set of $q$ that satisfies Eq.~\eqref{eq:domregion2}. Observe that $Z_{p'}^p$ is a hyper-rectangle with its axes parallel to the coordinate axes, starting at $m_{p', p}$ and going to infinity or negative infinity in the direction of $p'$.

Finally, note that $D_{p'}^p$ is a subset of $Z_{p'}^p$ that also satisfies Eq.~\eqref{ineq:exists}. The interior of $Z_{p'}^p$ always satisfies (\ref{ineq:exists}), but some points on the boundary of $Z_{p'}^p$ may not. E.g., $m_{p, p'}$ does not satisfy (\ref{ineq:exists}) by definition. Hence, $D_{p'}^p$ is the set $Z$ except some of its boundaries. Since $Z_{p'}^p$ is defined by a set of hyper-planes, to define $D_{p'}^p$, we also use the set of hyper-planes, but in addition to its coordinates we store the possible exceptions in the boundaries. We formalize our notation of hyper-planes later in this section.
Fig.~\ref{fig:DomRegions}(b)-(e) shows the domination regions of all points with respect to the $p_4$. 

\textbf{Skyline region of a point}. Recall that $S_p$ is the space where $p$ is not dominated by any other point. Using the terminology above, $S_p$ is the entire space except $\cup_{p'\in D}D_{p'}^{p}$, that is, $S_{p} = R^d\setminus(\cup_{p'\in D}D_{p'}^{p})$. This is because $\cup_{p'\in D}D_{p'}^{p}$ is the region where $p$ is dominated by some point in $D$. $S_p$ can be defined as a subset of $R^d$ except the union of hyper-rectangles with their axes parallel to the coordinate axes. 

Fig.~\ref{fig:SkyRegions}(d) shows how we can find the skyline region of point $p_4$. We first find the domination regions of all points with respect to $p_4$. Then we take their union and the skyline region of $p_4$ is the entire space except the union of the domination regions. As Fig.~\ref{fig:SkyRegions}(d) shows, not all points in $D$ contribute to the skyline region of $p_4$. %That is, skyline region of $p_4$ is the same irrespective of whether $p_3$ is in $D$ or not. This leads us to the concept of border points.

\textbf{Skyline hyper-planes}. The concepts defined so far consider subspaces of the query space defined by axis-parallel hyper-planes. In the rest of the paper, we refer by skyline hyper-plane (or simply hyper-plane) to the following data structure: 
for a hyper-plane $H$, the structure contains for each dimension $i$ two points $min[i]$ and $max[i]$, representing the smallest (largest) value on the hyper-plane in the $i$-th dimension. For any hyper-plane, there exists a dimension $i$ such that $min[i]=max[i]$. We say that a hyper-plane {\em is in} the $i$-th dimension if all the points on the hyper-plane have exactly the same value in that dimension. Furthermore, if the hyper-plane corresponds to the skyline region $S_p$, it stores the point $p$, and $p$ is referred to as the hyper-plane's {\em generator}. In general, $p$ is not a skyline point on the hyper-plane. However, as discussed before, a hyper-plane stores a set of exceptions, corresponding to coordinates (if any) on the hyper-plane for which $p$ is actually a skyline point. Furthermore, we say that a hyper-plane is bounded by a set of hyper-planes $H$ if every point on its boundary (defined by $min$ and $max$) also belongs to another hyper-plane in $H$.

\textbf{Border points}. Not all the points in $D$ contribute to $\cup_{p'\in D}D_{p'}^{p}$. That is, for two points, $p_1', p_2'\in D$, $D_{p_1'}^{p}$ may be a subset of $D_{p_2'}^{p}$, in which scenario $D_{p_1'}^{p}$ does not impact $S_p$. We refer to all the points such that $\forall p_2'\in D, D_{p_1'}^{p}\not\subseteq D_{p_2'}^{p}$ as \textit{border points} of $p$.  Fig. \ref{fig:SkyRegions} (a)-(e) shows the skyline regions of all points in the database. In each figure, the points in red are the border points and the points in blue are non-border points. The importance of border points is that they define how complex the skyline region of a point is. That is, the more number of border points, the more edges the skyline region will have. The following property helps us understand how many border points any point has. 

\begin{property}\label{Prop:borderSky}
$p$ divides the space into $2^d$ quadrants. Let $X_i$ contain the points in $D\setminus\{p\}$ that are in the $i$-th quadrant. $p'$ is a border point of $p$ if, for some $i$, $p'$ is a skyline point with respect to a query at $p$ for the database $X_i$.
\end{property}
%\noindent
%The number of border points is used in our analysis later on. 

\begin{figure*}[t]
    \begin{minipage}{0.32\textwidth}
        \centering
        \includegraphics[width=\textwidth]{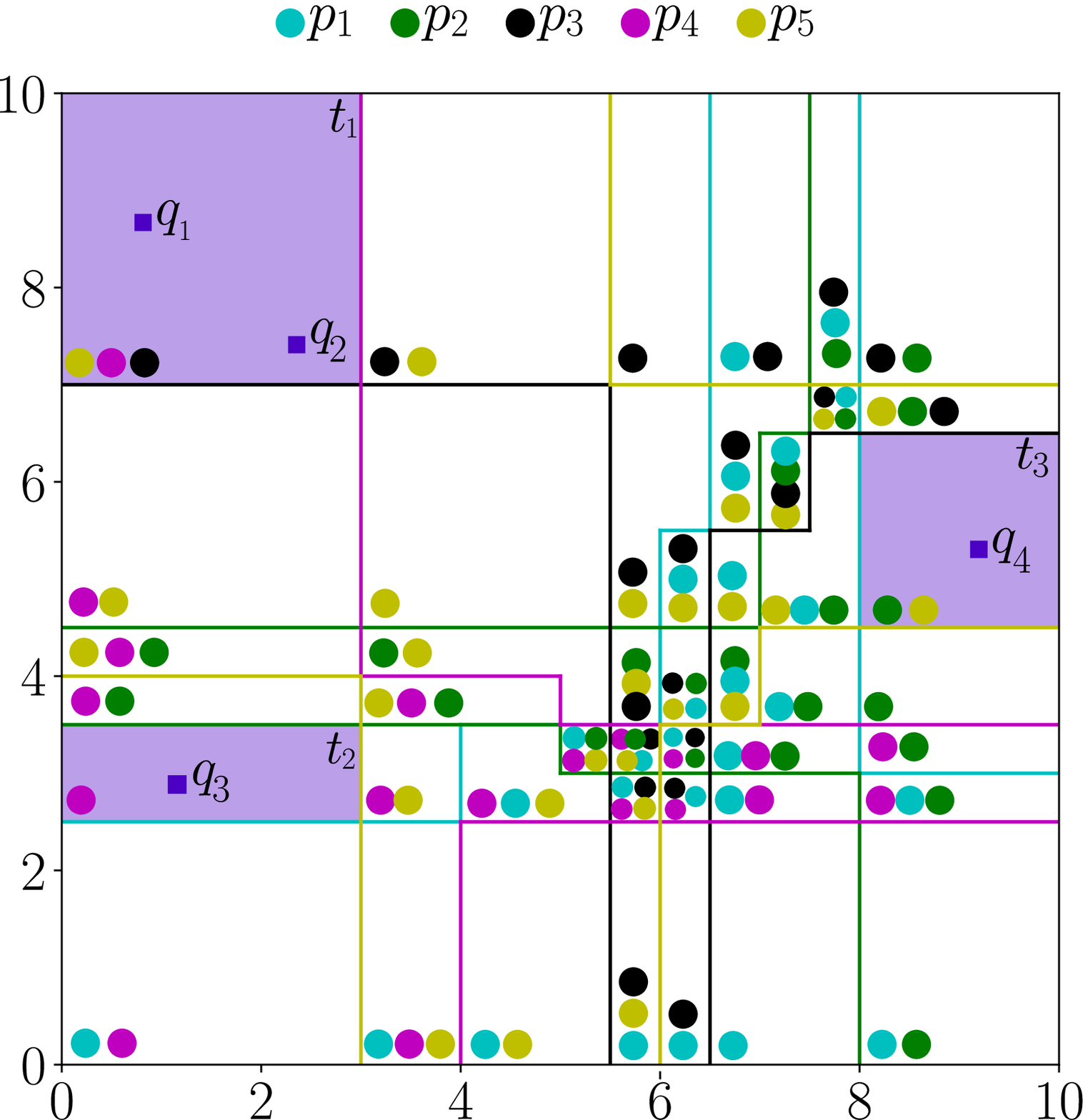}
        \vspace{-15pt}
        \caption{Skyline tiles}
        %\caption{Space tiling. Coloured circles show tile contents.}
        \label{fig:Tiles}
    \end{minipage}
    \hfill
    \begin{minipage}{0.32\textwidth}
        \centering
        \includegraphics[width=\textwidth]{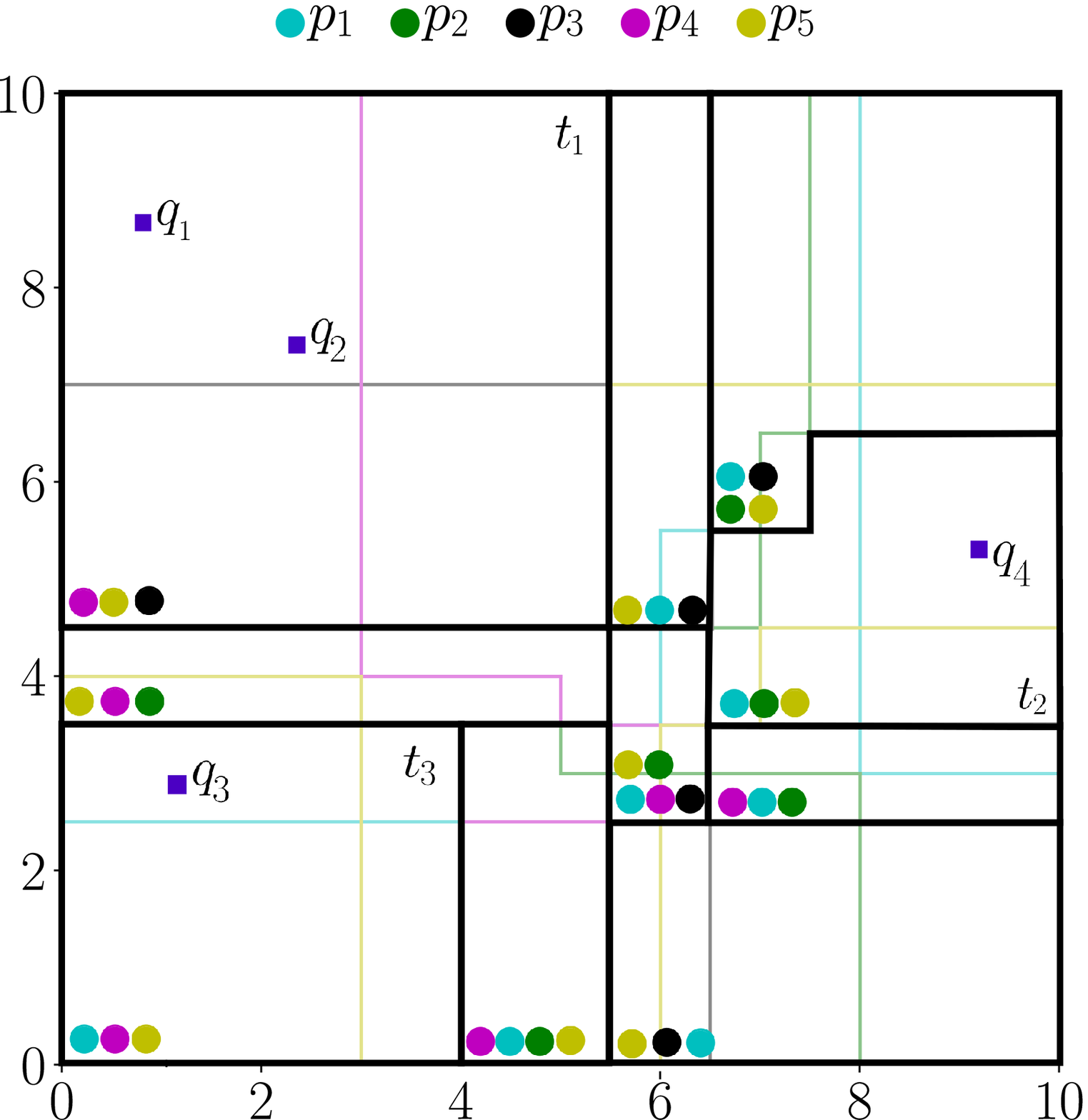}
        \vspace{-15pt}
        \caption{A solution to TAP}
        \label{fig:agg}
    \end{minipage}
    \hfill 
    \begin{minipage}{0.34\textwidth}
    \centering
    \includegraphics[width=\textwidth]{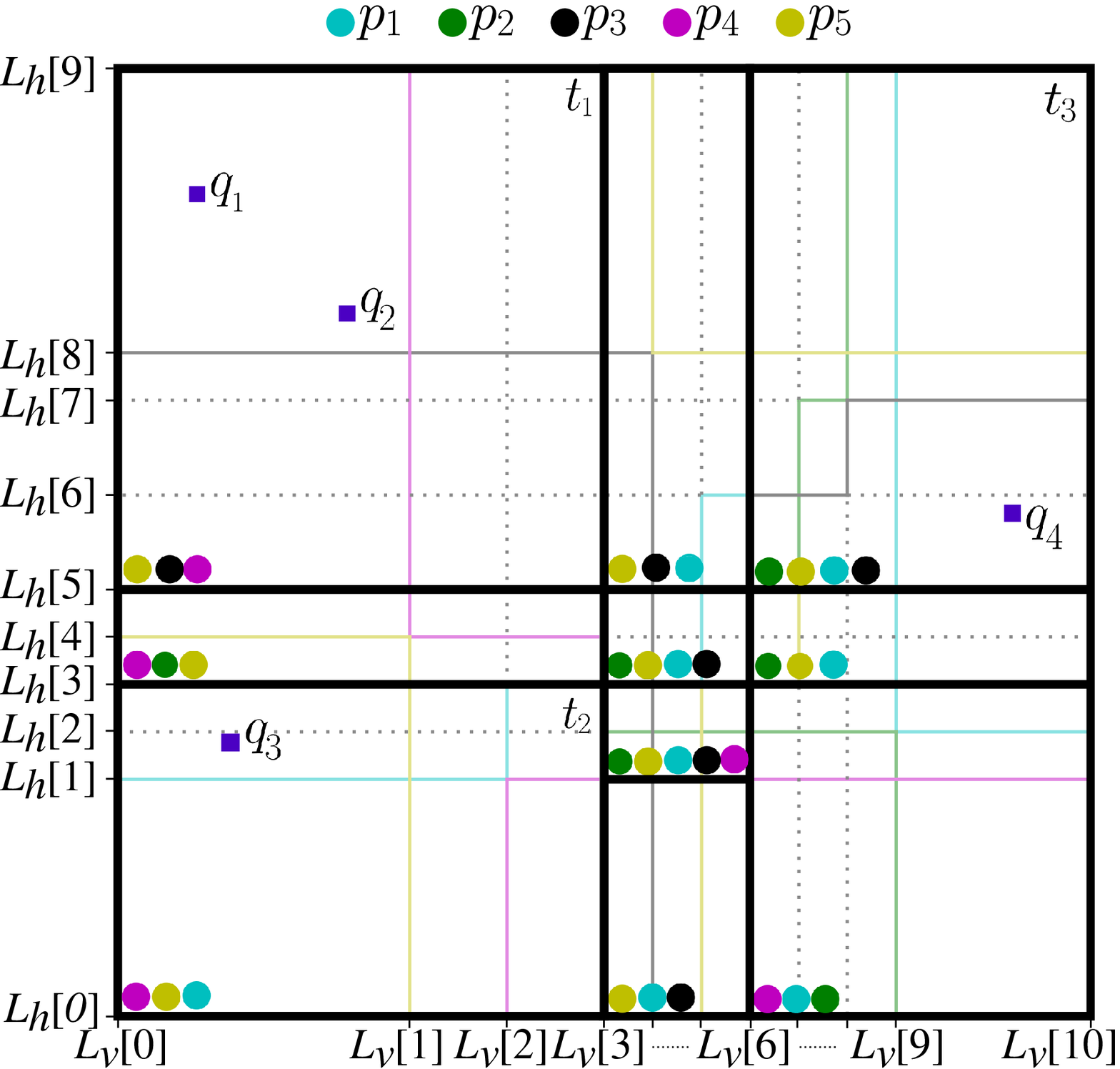}
    \vspace{-15pt}
    \caption{A solution to APP}
    \label{fig:partition}
\end{minipage}
\vspace{-15pt}
\end{figure*}

\subsection{Tiling of the Space}
%Next, we describe how we use the skyline region to partition the space into \textit{skyline tiles} that are used to answer the skyline queries. 

\subsubsection{Skyline Tiles}
Observe that for a query $q$ we can tell that $p$ is in the skyline iff $q\in S_p$. The answer $S$ to the skyline query $q$ is $S = \{p\in D|q\in S_p\}$. We need to find an efficient way to check whether $q$ is in $S_p$ for all points $p$ in $D$. To that end, we \textit{intersect} $S_p$ for all $p$ with each other. 

First, define a \textit{partitioning} of a space $Q$ as a set $\Pi$, such that for each $\pi\in \Pi$, $\pi\subseteq Q$, $\cup_{\pi\in \Pi}\pi=Q$ and that $\pi_i\cap\pi_j=\emptyset$ for any $\pi_i, \pi_j\in \Pi$, $\pi_i\neq\pi_j$. Let $H_p$ be the set of hyper-planes defining the skyline region, $S_p$, for a given $p$. Now consider the set $H=\cup_{p\in D}H_p$. Observe that $H_p$ is a set of intersecting hyper-planes where each hyper-plane is bounded by other hyper-planes. Consider the set of all polytopes created by the intersection of hyper-planes in $H_p$. Each polytope creates a partition, and their union $\Pi_D$ is a partitioning of the space. For the \textit{exceptions} stored in each hyper-plane, we also consider them to be a partition in the partitioning $\Pi_D$ (For ease of discussion, in the remainder of this paper, we do not explicitly mention the partitions created by the exceptions as their shape is different from the polytope partitions, but the discussion either directly holds for both or the extension to the exceptions is straight forward). We use this partitioning to define \textit{skyline tiles}. Formally, a \textit{tile} is defined as follows.

\begin{definition}[Tiles]
A \textit{tile}, $T$, is defined as a tuple $T=(S, P)$, where $S$ is a subspace of $R^d$ defined by a set of hyper-planes and $P$ is a subset of $D$. $S$ satisfies the following conditions. Firstly, each hyper-plane of $S$ is parallel to one of the axes. Secondly, each hyper-plane is bounded by another hyper-plane on all sides. 
\end{definition}

For a tile $T=(S, P)$, $S$ is called the \textit{location} of $T$ and $P$ is called the \textit{content} of $T$, also written as $cnt(T)=P$. Moreover, define \textit{space} of $T$, written as $spc(T)$, as the subset of $R^d$ that is inside $S$. We say a query \textit{falls} inside $T$ if $q\in spc(T)$. Finally, a \textit{tiling of the space} is a set of tiles, $X$, such that $\cup_{T\in X}spc(T)=R^d$ and $spc(T)\cap spc(T')=\emptyset$ for all $T, T'\in X$. 

For each partition $\pi_i\in\Pi_D$, a tile is $\tau_i$ defined as follows. Let $P_{\tau_i} = \{p\in D|\pi_i\cap S_p\neq\emptyset\}$ (i.e., the set of points whose skyline region intersects $\pi$). Let tile $\tau_i = (\pi_i, P_{\tau_i})$. We call $\tau$ a skyline tile and the set $\mathcal{T}_D=\{\tau_i, \forall i\}$ the set of skyline tiles of $D$. Skyline tiles have the following properties.

\begin{property}\label{prop:query_answer}
For a query $q$ that falls inside a skyline tile $T$, the answer to the skyline query at $q$ is $cnt(T)$.
\end{property}

\begin{property}\label{prop:tile_neighbour}
Consider two skyline tiles $T_1$ and $T_2$ such that both have a hyper-plane, $H$, as one of their edges. Assume hyper-plane $H$ correspond to some point $p$. Then, either $p\in cnt(T_1)$ or $p\in cnt(T_2)$ but not both. Furthermore, $cnt(T_2)\setminus\{p\}=cnt(T_1)\setminus\{p\}$. In other words, content of $T_1$ and $T_2$ differ in exactly one point, $p$.
\end{property}

Fig. \ref{fig:Tiles} shows the skyline tiles created from intersecting all the skyline regions in Fig. \ref{fig:SkyRegions}. In Fig. \ref{fig:Tiles}, observe that $t_1$ is a tile defined by four hyper-planes (or lines, since $d=2$), and its content is the set of points $\{p_3, p_4, p_5\}$. Now consider queries $q_1$ and $q_2$ in Fig. \ref{fig:Tiles}. Both queries fall in the tile $t_1$ and therefore their answer is $p_3$, $p_4$ and $p_5$. The query $q_3$ is in tile $t_2$ and its answer is $p_4$; whereas the answer to $q_4$ is $\{p_2,p_4\}$. This shows how an answer to a query can easily be retrieved by finding which tile the query falls into.

\textbf{Border Locations.} Recall that skyline tiles are created based on a set of hyper-planes, $H$. Consider the set $H_i$ of all the hyper-planes in $H$ that are in the $i$-th dimension. Let the set $L_i = \{x|h\in H_i,\; x=h.min[i]=h.max[i]\}$ (note that $h.min[i]=h.max[i]$ holds because $h$ is a hyper-plane in the $i$-th dimension). Observe that the set $L_i$ contains the $i$-th dimension boundary of all the skyline tiles. Define $N_i = |L_i|$ for all $i$ and let $N=\max_i{L_i}$. Observe that there is a relationship between number of border points and number of border locations. More specifically, every border point creates at most one hyper-plane in every dimension. Therefore, for every border point, we have a hyper-plane in $H_i$ that corresponds to a location in $L_i$. Thus, we can study the value of $N$ by analyzing the number of border points. This analysis is used in studying the performance of our algorithms. 

\vspace{-5pt}
\subsubsection{Analysis}
Two important properties of skyline tiles that are utilized in our analysis are the value of $N$ and the total number of tiles. They determine the space and time complexity of the algorithms discussed in the rest of the paper.

\textbf{Number of tiles.} Tiles are the intersection of $n$ different skyline regions, and each region contains at most $n$ hyper-planes in each dimension. For a hyper-plane in the first dimension, consider the maximum number of tiles it can be part of. In any other dimension, there can be at most $2^d\times n$ hyper-planes intersecting it . Thus, it can be a part of at most $2^{d(d-1)}n^{d-1}$ tiles. There are at most $n^2$ hyper-planes in the first dimension, and every tile must have one of those as its edge. Thus, there are at most $O(2^{d^2}n^{d+1})$ tiles in total.

\textbf{Value of $N$}. Let $B_p$ be the set of border points for $p$. According to Property \ref{Prop:borderSky}, $|B_p|$ is the size of a skyline query at $p$. Let $B_{avg} = \sum_{p\in D}\frac{|B_p|}{n}$. Observe that $N$ is at most $\sum_{p\in D}|B_p|$. Thus, we can write $N=n B_{avg}$. If data points are uniformly distributed, there will be $O(\frac{n(2\log n)^d}{d!})$ number of skyline points~\cite{buchta1989average} ($\frac{(\log n)^d}{d!}$ is the expected number of skyline points for uniformly distribution, there are $n$ data points, and we need to consider skyline points for each of the $2^d$ quadrants created by $p$). Therefore, in such a scenario, $B_{avg}=O(\frac{n(2\log n)^d}{d!})$ on expectation. 

\textbf{Challenges}. The number of tiles created by this approach is exponential in data size and makes the problem intractable. This occurs because
% is due to the lack of locality of the skyline regions. That is, 
skyline regions may impact parts of the space far from their generating point.
% (contrast with Voronoi diagrams that are more localized). 
Working directly with skyline tiles may be inefficient. We present two different methods in Sections~\ref{sec:agg} and~\ref{sec:genTiles} to deal with this issue.

%% file: TileAggregation.tex
\section{Aggregating Tiles}\label{sec:agg}
%Although the approach introduced in Section \ref{sec:tiling} allows us to efficiently answer skyline queries, 
One approach to address the high space complexity of skyline tiles is to aggregate some of the tiles together. This saves spaces by storing the content of multiple tiles only once, but it may require the inclusion of false positives in the answer. Recent work~\cite{SNN} utilizes the computational power of the user to filter the final results based on the data returned to the user. 
%That is, we can return a super-set of the answer to the user and perform fine tuning using the user's computational power to obtain the final answer. 
Considering that the user can decrypt the data and operate on plaintexts, the amount of work required to filter out the false positives is very small. 
%Note that since the fine tuning at user side will be using plane-text operations (i.e., not performed on encrypted data), the operation can be very fast and the cost of operations at users side will still be dominated by the decryption operations which take much longer than the fine tuning step \textbf{[evidence]}. 
On the other hand, allowing a small amount of false positives can significantly increase storage and processing efficiency at the SP. 
%We allow returning super-sets of the skyline query to reduce the space requirement of our method. 
In essence, we no longer partition the data domain so that the answer to each skyline query \textit{is the same} within each tile. Instead, we partition the domain such that the answer to a query \textit{does not change by much} within each partition. We achieve this by combining some of the tiles together. The tile aggregation problem is formalized as follows.

\begin{figure}
%    \begin{minipage}{0.66\textwidth}
        \centering
        \includegraphics[width=\columnwidth]{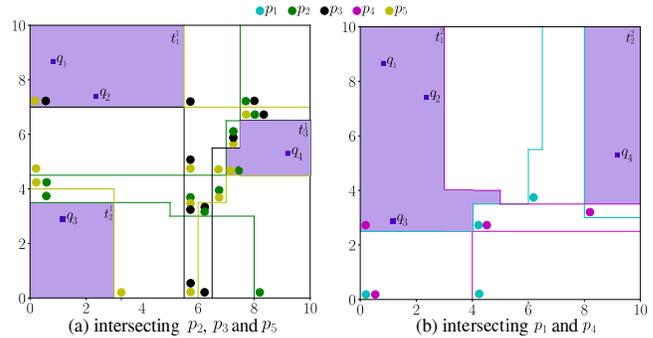}
        \vspace{-20pt}
        \caption{Generalized tiling with $l=3$.}
        \label{fig:MultiTiles}
\vspace{-15pt}
%    \end{minipage}
\end{figure}

\subsection{Tile Aggregation Problem}
When false positives are permitted, a solution $S^q$ returned by skyline processing algorithm for query $q$ consists of two sets: set $S_c^q$ which contains only (and all) points that are in the skyline of $q$, and the set $S_f^q$ which contains none of the skyline points of $q$. We refer to the points in $S_f^q$ as \textit{false hits} and $|S_f^q|$ as the number of false hits for the query $q$. 

For efficient post-processing, we bound the number of false hits for any query. We define the \textit{false-hit requirement} as follows: let $k$ be an integer; then the maximum number of false hits in a solution for any query must be at most $k$, that is $\max_q |S_f^q|\leq k$. Recall that, by allowing false-hits, we aim at reducing the space complexity of our skyline result materialization structure. We reduce the space complexity by \textit{aggregating} some of the skyline tiles. An aggregation is formally defined as follows: let $T =\{t_1=(S_1, P_1), t_2=(S_2, P_2), ..., t_r=(S_r, P_r)\}$ for the following definitions.
 
\begin{definition}[Aggregation]
An \textit{aggregation} (or \textit{aggregate tile}), $A$, of a set of tiles $T$ is itself a tile $A=(S, P)$ that satisfies the following conditions. (1) $P= \cup_iP_i$ and (2) $S$ is the smallest set such that $\cup_{i}S_i\subseteq S$. The tiles in $T$ are called the \textit{component tiles} of $A$. 
\end{definition}

\begin{definition}[Location-wise validity]
An aggregation $A$ of $T$ is \textit{location-wise valid} if the aggregation creates a single connected polytope. 
\end{definition}

Intuitively, an aggregation is location-wise valid if all of its component tiles are next to each other, i.e., there is no empty space between them. 

\begin{definition}[Cardinality-wise validity]
We say that an aggregation $A$ of $T$ is \textit{cardinality-wise valid} if $|P|\leq k+\min_r|P_i|$, for a parameter $k$. 
\end{definition}

That is, the aggregation contains at most $k$ additional points compared to any of its component tiles. 
Observe that if an aggregation $A$ is location-wise valid, any query $q$ that falls inside the aggregation also falls inside one of its component tiles, $T$. If the aggregation is also cardinality-wise valid, we can return the content of $A$ instead of $T$ to answer the query $q$, and the solution will satisfy the false-hit requirement. Thus, we aim at finding aggregations that are both location-wise and cardinality-wise valid. Furthermore, we aim to reduce the total space used by the algorithm. We can express this as an optimization problem as follows. 

\begin{definition}[Tile Aggregation Problem]
Given a set of tiles $T$ and an integer $k$, return set $S\subseteq 2^{T}$ such that $\cup_{t\in S}=T$, $\forall t_i\in S$ the aggregation $a_i$ of $t_i$ is both location-wise and cardinality-wise valid, and $\sum_{i}|cnt(a_i)|$ is minimized.  
\end{definition}

Fig.~\ref{fig:agg} shows a feasible solution to the Tile Aggregation Problem (TAP) when $k=2$. Observe that the content of each aggregated tile is the union of the points in each of the component tiles. Furthermore, for queries $q_1$ and $q_2$, the answer is the same as the answer with no aggregation (see Fig.~\ref{sec:tiling}). However, $q_3$ now returns two false hits, i.e., $p_1$ and $p_5$ while $q_4$ returns one false hit, i.e., $p_1$. 

The tile aggregation problem (TAP) as defined above is difficult to solve optimally. Specifically, we show that it is NP-hard even in two dimensions.

\begin{theorem}\label{thm:NPHard}
TAP is NP-hard.
\end{theorem}

\textit{Proof.} See Appendix~\ref{sec:appx}. %\textbf{Mention that it's not for Skyline but TAP}

Two issues arise when solving TAP. First, solving TAP optimally is NP-hard. Second, the aggregate tiles of the TAP solution can have complicated shapes, slowing down the process of searching them. We address these issues next. %However, observing the possible solutions to TAP allows us to formulate the problem differently.

\subsection{Relaxed Aggregations}\label{sec:partitioning}
We propose a relaxation of the aggregation problem. We restrict the possible choices by enforcing that aggregations must have a certain shape. However, we allow aggregations to split existing tiles into two (that is, half of a tile may belong to one aggregations and the other half to another) to avoid over-restrictive  requirements. These modifications make it more intuitive to formulate the problem as a space partitioning problem, as discussed below.

Note that, any feasible solution, $S$, to TAP corresponds to a set of aggregations $A$ whose union of space, i.e., $\cup_{a\in A} spc(a)$, covers the entire data domain. On the other hand, if we allow splitting of the tiles during aggregation, the observation here is that in fact \textit{any} partitioning of the space, can be used to create an aggregation that in turn can be used to answer the skyline query. Consider a partitioning of the query space $\Pi=\{\pi_1, \pi_2, ..., \pi_r\}$. Define a set of tile $T=\{t_1, ...,. t_r\}$ such that $spc(t_i)=\pi_i$. Moreover, for a tile $t_i$, let $B_i=\{\tau\in \mathcal{T}_D|spc(\tau)\cap spc(t_i)\neq \emptyset\}$ and set $cnt(t_i)=\cup_{\tau\in B_i}cnt(\tau)$. Note that the set $T$ defines a tiling of the space that covers the entire domain and if a query falls into a tile, the content of the tile will be a super-set of the answer to the query.

Since every partitioning of the query space corresponds to a tiling by the construction above, we formulate this relaxation of TAP (this is a relaxation because we now allow splitting of tiles), in terms of a space partitioning problem that minimizes the storage cost. Finding a space partitioning in the general case is also difficult, since, many aggregations are also space partitioning. %TODO:\textbf{Optimal space partitioning same as TAP}
However, we reduce the complexity of the problem by restricting the shape of a partitioning allowed. 

\begin{definition}[Shape-wise validity]
A partitioning is \textit{shape-wise valid} if it can be represented by a set of hyper-planes where a hyper-plane in the $i$-th dimension has boundaries from $-\infty$ to $\infty$ in all dimensions $j$ when $j>i$. 
\end{definition}
Intuitively, the partitioning contains hyper-planes in dimension $i$ that are not bounded by any hyper-plane in dimensions after $i$. 
%TODO:\textbf{ADD a figure, and benefits, define min and max val, tiling and partitioning interchangble? Maybe change aggregation to some other word}

\begin{definition}[Aggregation by Partitioning Problem]
For the query space $Q$, find a set of hyper-planes that define a shape-wise valid partitioning of $Q$ such that the tiling, $T$, corresponding to the partitioning is cardinality-wise valid and that $\sum_{t\in T}|cnt(t)|$ is minimized.
\end{definition}

%\textbf{Need to say why we bring up query space. Maybe put that in the definition of partitioning}. 
We provide a dynamic programming solution that solves Aggregation by Partitioning Problem (APP) optimally and then discuss a number of heuristics that solve it without approximation guarantees. It is worth mentioning that the quality of our solution to APP (i.e., whether it is optimal or not), helps with reducing storage cost of our index. However, our guarantees regarding properties of tiles (e.g., the false-hit requirement) hold true irrespective of whether the problem is solved optimally or not. Thus, heuristics can be useful in practice when using minimal space is not critical.

\subsubsection{Dynamic Programming Solution to APP}\label{sec:DP}
We can solve APP optimally by dynamic programming. We limit our discussion to two-dimensions for ease of illustration. The idea can be extended to higher dimensions, and we can obtain a polynomial running time for any fixed dimensionality. However, in practice, the run time will become large for high dimensions, and using heuristics may be a better option in those scenarios.

The dynamic programming formulation uses two observations. First, the shape-wise validity requirement of APP in two dimensions implies the that the space partitioning is defined by a set of vertical lines that cross the entire space and a set of horizontal lines that start and end between adjacent vertical lines. Second, there exists an optimal solution where all line overlaps the boundary of some skyline tile. This is because for any line that does not overlap the boundary of any skyline tile, we can move that line until it does overlap the boundary, and the total cost will not increase. Thus, to find the optimal solution, we only need to consider lines that pass through the boundaries of the skyline tiles.

\textbf{Recurrence relation.} Let $L_v$ be an array of all possible $x$ locations for the vertical lines and let $L_h$ be an array of all possible $y$ locations for the horizontal lines, both sorted in ascending order. To obtain $L_v$ and $L_h$, we enumerate all the skyline tiles and find the $x$ and $y$ values of their boundary lines and add them to $L_v$ and $L_h$ if they don't exist. Let $N_h=|L_h|$ and $N_v=|L_v|$. Furthermore, consider $R=((x_1, y_1), (x_2, y_2))$ as a rectangle with lower left coordinates $(x_1, y_1)$ and upper right coordinates $(x_2, y_2)$. Define $C(s, i, t, j)$ as the size of the content of the aggregate tile whose space is $R=((L_v[s], L_h[i]), (L_v[t], L_h[j]))$ or infinity if such an aggregation is not cardinality-wise valid. That is, let $B=\{\tau\in \mathcal{T}_D|spc(\tau)\cap R\neq \emptyset\}$. Define $smallest=\min_{\tau\in B}|cnt(\tau)|$ and $size=|\cup_{\tau\in B}cnt(\tau)|$. Then let
\vspace{-8pt}
$$
C(s, i, t, j)=\left\{\begin{matrix}
 size & smallest+k\leq size \\ 
 \infty & \text{otherwise}
 \end{matrix}\right.
\vspace{-8pt}
$$
Define $V(i)$ as the optimal solution to the problem of APP in the space where $x>L_v[i]$, given that there exists a vertical line at $L_v[i]$. Note that the optimal solution to our problem is $V(0)$. Furthermore, define $H(i; s, t)$ as the optimal solution to APP in the space $L_v[s]<x<L_v[t]$, $y>L_h[i]$ with only horizontal lines given that there are vertical lines at $L_v[s]$ and $L_v[t]$ and a horizontal line at $L_h[i]$. Then, We can write the following recurrence relations.
\vspace{-7pt}
\begin{align}
V(i) = \min_{i<j\leq N_v} H(0; i, j)+V(j)\\
H(i; s, t) = \min_{i<j\leq N_h} C(s, i, t, j) + H(j; s, t)
\end{align}
\noindent
with the base cases $H(N_h; s, t)=0$ for all $s, t$ and $V(N_v)=0$. Intuitively, the recurrence relation tries to find the next vertical line after a given location and breaks down the problem into two separate instances of smaller size: (1) from the current line to the next and (2) from the next line to the end of the space. Doing so is possible because the vertical lines go through the entire space, that is the instance of the problem before a vertical line is independent of the instance after the vertical line. Furthermore, instance (1) contains only horizontal lines, therefore, we can solve it recursively by only considering all possible horizontal lines.

\begin{algorithm}[t]
\begin{algorithmic}[1]
\STATE $L_h\leftarrow$ horizontal boundaries of $\mathcal{T}_D$, sorted
\STATE $L_v\leftarrow$ vertical boundaries of $\mathcal{T}_D$, sorted
\STATE $N_v\leftarrow |L_v|$ and  $N_h\leftarrow |L_h|$
\STATE $V(N_v)\leftarrow 0$
\FOR{$i\leftarrow N_v-1$ \TO $0$}
    \STATE $V(i)\leftarrow \infty$
    \FOR{$j\leftarrow i+1$ \TO $N_v$ \AND $j-i\leq v_{max}$}
        \STATE $H(N_h)\leftarrow 0$
        \FOR{$s\leftarrow N_h-1$ \TO $0$}
            \STATE $H(s)\leftarrow \infty$
            \FOR{$t\leftarrow s+1$ \TO $N_h$  \AND $t-s\leq h_{max}$}
                \STATE $c \leftarrow C(i, s, j, t)+H(t)$
                \IF{$c\leq H(s)$}
                    \STATE $H(s) \leftarrow c$
                \ENDIF
            \ENDFOR
        \ENDFOR
        \STATE $c \leftarrow H(0)+V(j)$
        \IF{$c\leq V(i)$}
            \STATE $V(i) \leftarrow c$
        \ENDIF
    \ENDFOR
\ENDFOR
\RETURN V(0)

\caption{DP($\mathcal{T}_D$)}
\label{algo:DP}
\end{algorithmic}
\end{algorithm}

\textbf{Algorithm.} Algorithm~\ref{algo:DP} implements the recurrence relation. The algorithm starts backwards and tabulates the values for $H$ and $V$. Note that one optimization compared with the recurrence relation is that distance between $i$ and $j$ is bounded by $v_{max}$ (and similarly $s$ and $t$ by $h_{max}$). This is because of the false-hit requirement of the solution. That is, $v_{max}$ is the maximum possible integer such that the space from $L_v[v_{max}]$ to $L_v[v_{max}+i]$ has a feasible solution using only horizontal lines (i.e., a partitioning exists that satisfies the false-hit requirement) for all $i\leq |L_v|-v_{max}$. 

\textbf{Correctness.} Observe that, given that a vertical line exists at $L_v[i]$, the optimal solution must contain a next vertical line at a location $L_v[j]$, for some $j>i$ (except for the base case). Each $j$ splits the space into two: (1) the space from $L_v[i]$ to $L_v[j]$; and (2) the space after $L_v[j]$. Note that, no partition is allowed to cross $L_v[j]$ because the partitioning has to be shape-wise valid. As a result, the two instances can be solved independently. Consider instance (1): it is the problem of finding the optimal tiling of the space between  $L_v[i]$ and $L_v[j]$ using only horizontal lines, given that there are vertical lines at $L_v[i]$ and $L_v[j]$. This is the same as $H(0; i, j)$. Consider instance (2): it is the problem of finding the optimal tiling of the space after $L_v[j]$ given that there is a vertical line at $L_v[j]$. This is the same as $V(j)$. Therefore, for each $j$, the minimum possible cost is $H(0; i, j)+V(j)$. Since the optimal solution has to contain one of the possible values of $j$, then its cost must be $\min_j H(0; i, j)+V(j)$. A similar argument proves the correctness of the recurrence relation for $H$. 

%TODO:\textbf{Multiple set of tiles, analysis including $l$}

\textbf{Time Complexity.} As Algorithm \ref{algo:DP} shows, there are four nested loops, and each of them take $N_v$, $v_{max}$, $N_h$ and $h_{max}$ respectively. $C(i, s, j, t)$ can be found in $\log n$ using Property \ref{prop:tile_neighbour} (we keep track of the content of the tiles, and whenever a hyper-plan is crossed, we use Property \ref{prop:tile_neighbour} to determine the content of the new tile encountered). $O(\log n)$ is needed to determine whether the differing point mentioned in Property \ref{prop:tile_neighbour} already exists in the previous tile or not. Let $N=\max\{N_v, N_h\}$. Thus, the total running time is $O(v_{max}h_{max}N^2\log n)$ ($C(i, s, j, t)$ can also be pre-computed to avoid the $\log n$ factor at the expense of storage cost). $v_{max}$ and $h_{max}$ depend on $k$ and on how the tiles are distributed. They are generally similar to $k$ in value, since every skyline tile differs in exactly one point from any of its neighbors. 

\subsubsection{Heuristics to Reduce Computation Time}\label{sec:heuristics}
We present two heuristics that can be used either independently, or together with our dynamic programming approach, to improve the running time of our algorithm at the cost of losing optimality. 

\textbf{Prepartitioning}. To reduce the time complexity of the algorithm, we first \textit{pre-partition} the space. We can do this by placing a vertical and horizontal line at every every multiple of $m$ coordinates, where $m$ is a fixed parameter. This bounds $v_{max}$ and $h_{max}$ to $m$. We solve the problem for each partition created using DP. Each partition now has $m$ possible vertical positions and $m$ horizontal positions.  The run-time of the algorithm is reduced to $O(N^2m^2\log n)$, where $m$ can be used to trade-off optimality with run-time. The solution approaches the optimal as $m$ increases.

\textbf{Greedy.} Another heuristic is to choose the lines greedily, that is, starting from the beginning and choosing the next vertical line as far away from the current position as possible, and repeating that for the horizontal dimension. This reduces the cost to $O(|\mathcal{T}_D|)$. %factor that is used to solve the vertical problem to $O(\frac{N}{v_max})$. Note that with this approach, increasing $k$ improves the running time as opposed to worsening it, which maybe useful if we want to try larger values of $k$.

%% file: GenTiling.tex
\section{Generalized Skyline Tiles}\label{sec:genTiles}
Another way to reduce the storage overhead is to generalize the skyline tile concept, and to allow for a tunable parameter providing a trade-off between space complexity and query time.
%I eliminated the subsection, since it is the same as the section title, so not much point to consume the extra space
%\subsection{Generalized Skyline Tiles}
Recall that, to generate skyline tiles, we intersected the skyline region of \textit{all} the points, which lead to the creation of a large number of tiles. In this section, we generalize the process by only intersection $l$ of the skyline regions, for a parameter $l$ which gives us $\lceil\frac{n}{l}\rceil$ different sets of tiles. This helps reducing the space complexity, because it reduces the fragmentation of tiles (fewer data points result in fewer skyline region intersections). However, it increases query time, since we need to search multiple sets of tiles to find the final answer to the query. In the extreme case when $l = n$, we obtain one set of tiles that intersects all the skyline regions. When $l=1$ we do not intersect the skyline regions at all, but we use each skyline region individually to know whether a point is in the skyline or not.

Observe that, generalizations of Property~\ref{prop:query_answer} and Property~\ref{prop:tile_neighbour} hold for generalized skyline tiles. For Property~\ref{prop:query_answer}, observe that a skyline tile created from the intersection of a set of points $D_i\subseteq D$ contains the subset of the answer to the skyline query in $D$. In other words, the union of the tiles a skyline query falls into is the answer to the skyline query.
Furthermore, note that the aggregation methods discussed in Section~\ref{sec:agg} can still be applied to each set of generalized tiles created. That is, we apply the method $\lceil\frac{n}{l}\rceil$ times. However, the total number of false-hits will now be $\lceil\frac{n}{l}\rceil\times k$, as $k$ is the number of false hits per use of the aggregation method.

Fig.~\ref{fig:MultiTiles} illustrates the generalization concept. Setting $l=3$, we get two different sets of tiles.  Fig.~\ref{fig:MultiTiles}(a) shows the intersection of skyline regions for $p_2$, $p_4$ and $p_5$, whereas Fig.~\ref{fig:MultiTiles}(b) shows the intersection for the remaining skyline regions. Note that, a query has to search both sets of tiles. For instance, queries $q_1$ and $q_2$ fall into the tile $t_1^1$ from which we obtain points $p_5$ and $p_3$. They also fall into the tile $t_1^2$ from which we obtain $p_4$. Thus, the answer to $q_1$ and $q_2$ is $p_3$, $p_4$ and $p_5$. Observe that $q_3$ falls into $t_2^1$, but $t_2^1$ now does not contain any point at all (this is possible when the intersection is not done on all the skyline regions). $q_3$ also falls into $t_1^2$ which contains $p_4$. Thus, the answer to $q_3$ is $p_4$. 

\subsection{Analysis}
\textbf{Number of Tiles}.
Each index contains $l$ skyline regions and each skyline region contains at most $n$ hyperplanes in each of the $d$ dimensions. For a hyperplane in the first dimension, consider the maximum number of tiles it can be a part of. There can be at most $2^d\times l$ hyper-planes intersecting it in any other dimension. Thus, it can be a part of at most $2^{d^2}l^{d-1}$ tiles. There are at most $nl$ hyperplanes in the first dimension, and every tile has to have one of those as its edge. Thus, there are at most $O(2^{d^2}nl^{d})$ tiles in total for each set of tiles. %\textbf{Mention kd-tree takes more space, we could do other indices at the expence of query time}.

\textbf{Value of $N$}. Let $B_p$ be the set of border points for $p$. According to Property~\ref{Prop:borderSky}, $|B_p|$ is the size of a skyline query at $p$. Let $B_{avg} = \sum_{p}\frac{|B_p|}{l}$. Since $N$ is at most $\sum_{p}|B_p|$, we can write $N=l B_{avg}$. If data points are uniformly distributed, there will be $O(\frac{n(2\log n)^d}{d!})$ skyline points. This is because $\frac{(\log n)^d}{d!}$ is the expected number of skyline points if the points are uniformly distributed, there are $n$ data points, and we need to consider the number of skyline points for each of the $2^d$ quadrants created based on $p$. Therefore, in such a scenario, $B_{avg}=O(\frac{n(2\log n)^d}{d!})$ on expectation. 

\textbf{Remark}. Observe that based on our choice of $l$, we can avoid both space and time complexity $n^d$ (for instance, by setting $l=n^{\frac{1}{d}}$). Thus, generalized tiling is particularly useful as dimensionality increases.

%% file: FinalAlgo.tex
\vspace{-10pt}
\section{Complete Skyline Algorithm}\label{sec:algo}

%\begin{algorithm}[t]
%\begin{algorithmic}[1]
%\STATE Need to decide indexing method to see if it's worth first creating all the tiles
%\caption{GetTiles($S$)}
%\label{algo:getTiles}
%\end{algorithmic}
%\end{algorithm}

\begin{figure}
    \centering
    \includegraphics[width=\columnwidth]{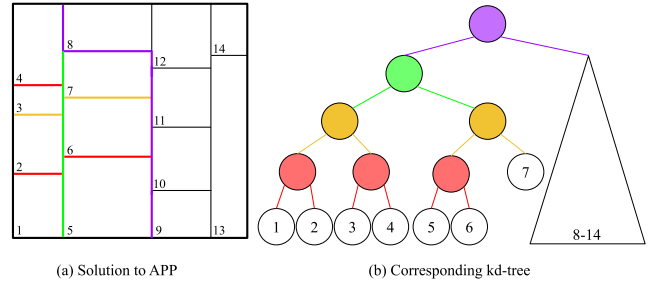}
    \vspace{-20pt}
    \caption{Building a kd-tree for a solution to APP}
    \vspace{-10pt}
    \label{fig:building_tree}
\end{figure}

\begin{algorithm}[t]
\begin{algorithmic}[1]
\STATE $B\leftarrow$ border points of $p$
\STATE $S_p\leftarrow\emptyset$
\FOR{$p_{i} \in B$}
    \STATE $m\leftarrow \frac{p+p_i}{2}$
    \FOR{$dim\leq d$}
        \STATE $h\leftarrow$ d-dimensional hyper-plane
        \FOR{$dim'\leq d$}
            \STATE $m\leftarrow \frac{p[dim']+p_i[dim']}{2}$
            \STATE $end \leftarrow$ end point of hyper-plane\label{alg:sky_endpoint}
            \STATE $h[dim]\leftarrow(m, end)$
        \ENDFOR
        \STATE $S_{p}\leftarrow S_{p}\cup \{h\}$
    \ENDFOR
\ENDFOR
\RETURN $S_p$
\caption{GetSkylineRegion($D$, $p$)}
\label{algo:getSkylineRegion}
\end{algorithmic}
%\vspace{-10pt}
\end{algorithm}

\begin{algorithm}[t]
\begin{algorithmic}[1]
\FOR{$p_{i} \in D$}
    \STATE $S_{p_i}\leftarrow$ GetSkylineRegion($D$, $p_i$)
\ENDFOR
\STATE $\mathcal{I}\leftarrow\emptyset$
\FOR{$i \leq \lceil\frac{n}{l}\rceil$}
    \STATE $H\leftarrow AggregateTiles(k, S_{p_{i\times l}}\cup S_{p_{i\times l+1}}...\cup S_{p_{(i+1)\times l}})$\label{alg:index:line:aggregate}
    \STATE $\mathcal{I}_i\leftarrow$ kd-tree on $H$  
    \FOR {\textbf{each} leaf node $n_l$ \textbf{in} $\mathcal{I}_i$}
        \STATE $n_l.points\leftarrow$ GetSkylinePoints($n_l$)
    \ENDFOR
    \STATE Add $\mathcal{I}_i$ to $\mathcal{I}$
\ENDFOR
\RETURN $\mathcal{I}$
\caption{GetSkylineIndices($D$, $l$, $k$)}
\label{algo:getIndex}
\end{algorithmic}
\end{algorithm}

\begin{comment}
\begin{algorithm}[t]
\begin{algorithmic}[1]
\STATE $S\leftarrow\emptyset$
\FOR{$\mathcal{I}_i\in \mathcal{I}$}
    \STATE $S\leftarrow S\cup \mathcal{I}_i.find(q)$
\ENDFOR
\RETURN $S$
\caption{Query($q$, $\mathcal{I}$)}
\label{algo:Query}
\end{algorithmic}
\end{algorithm}
\end{comment}

%\subsection{Algorithm and Analysis}\label{sec:algo_analysis}
We described several methods to pre-compute the skyline result for any query within the data domain. Our approaches achieve various trade-offs between computation time, storage size and number of false hits. %, by employing dynamic programming or greedy strategies in the tile aggregation process, or by partitioning the dataset and running the pre-computation on several different problem instances. 
In this section, we show how the final query can be determined using plaintext data and returned to the user based on our pre-computed set of results. In the next section, we present how the data structures are encrypted, and how the query answering operations are performed using encrypted data.

The performance of our approach can be tuned using two main parameters: $l$ and $k$. The former determines the number of partitions we divide our dataset into, and the latter the number of false positives allowed. Setting $k=0$, the set of tiles will provide an exact answer, while in general there can be $\lceil\frac{n}{l}\rceil k$ number of false hits in the answer. Following the result pre-computation, we generate a set of indices, $\mathcal{I}$, containing $\lceil\frac{n}{l}\rceil$ separate index structures, used to answer the skyline query. We call $\mathcal{I}_i\in\mathcal{I}$ a skyline index. 

%\subsection{Index Preprocessing}
According to our validity conditions (Sec.~\ref{sec:agg}), $\mathcal{I}_i$ must partition the space recursively while disallowing overlaps between partitions. We employ a {\em kd-tree} index for this purpose. Alg.~\ref{algo:getIndex} illustrates the combined process of result pre-computation and index construction, with four stages:
% {\em (1)} finding the hyper-planes for all skyline regions, {\em (2)} aggregating tiles, which returns a set of hyper-planes defining the aggregation, {\em (3)} indexing the hyper-planes in a kd-tree, and {\em (4)} assigning skyline points to each leaf node of the kd-tree. Next, we briefly outline each of these stages.

\textbf{1. Construction of Skyline Region.} 
%As Algorithm \ref{algo:getSkylineRegion} shows, 
To create the skyline region of a point $p$, we first find its border points by issuing a skyline query at $p$ on $D\setminus\{p\}$. Then, we iterate through all the border points, $p'$ of $p$ and find $D_{p'}^p$. $D_{p'}^p$ is defined in terms of a number of hyper-planes. Thus, we store these hyper-planes for each points for future use.

\textbf{2. Aggregating Tiles.} Line~\ref{alg:index:line:aggregate} of Alg.~\ref{algo:getIndex} uses one of the methods discussed in Sec.~\ref{sec:agg} to perform tile aggregation and returns the resulting hyper-planes. Note that, even when $k=0$, we run an APP algorithm with $k=0$ and obtain the corresponding hyper-planes. This approaches traverses the tiles once and splits some tiles into two, but constructs hyper-planes that can be easily used to create a balanced indexed (see below). Alternatively, when $k=0$, we can skip this step but the process of creating a balanced tree become more complicated.

\textbf{3. Building kd-tree.} 
%To allow for better security guarantees (See. Sec. \ref{sec:encrypteion}), 
We build a balanced kd-tree from our tiles, as follows. All the hyper-planes are given in advance by our solution to APP, so we build a balanced kd-tree. Since for a solution to APP, the hyper-planes in the $i$-th dimension do not cross hyper-planes in the $j$-th dimension for $j<i$, we impose the following ordering on tiles by utilizing the partitioning. We traverse the tiles by going through the hyper-planes in the first dimension iteratively, and for each hyper-plane, recursively going through the hyper-planes in the next dimensions that fall right before it. Fig. \ref{fig:building_tree} (a) shows how we can do this in two dimensions.  We start from the left-most vertical line, and go through the tile in the ascending order of the horizontal lines. Then we move on to the next vertical line. This gives us the ordering shown in Fig.~\ref{fig:building_tree}(a), where the numbers show the position of each tile in that order. 

Then, to build the tree, we choose the split points so that half of the tiles are stored in one sub-tree and the other half in another. For instance, in Fig.~\ref{fig:building_tree}, tiles 1-7 are in the left sub-tree and tiles 8-14 are in the right sub-tree. Note that the condition for the split can be define by $2\times d - 1 $ number of comparisons at each node. For instance, the condition for splitting at the root in Fig.~\ref{fig:building_tree}(b) is shown by the purple lines in Fig.~\ref{fig:building_tree}(a). It shows how two vertical lines and one horizontal line is enough to separate tiles 1-7 from tiles 8-14. This process is repeated recursively (each coloured line in Fig.~\ref{fig:building_tree} corresponds to the condition for a node with the same colour). Observe that the leaf nodes in the final tree created will have heights that differ by at most one. strictly speaking, the created index is not exactly a kd-tree, as the splitting conditions are different than a typical kd-tree.

%Given a set of hyper-planes, a kd-tree node $n_k$ splits the space across a dimension $i$ based on the median of $L_i$. The kd-tree recursively inserts the hyper-planes falling inside each of the half-spaces into the children of the node. We recursively split the children, but in a different dimension until no hyper-plane crosses a child. Observe that each leaf node now covers the space that is a subset of each (aggregate) tile.

\textbf{4. Assigning Skyline Points.} For each leaf node of the kd-tree we assign the corresponding skyline points. The content of each leaf node is already determined by running APP. Thus, we traverse all the leaf nodes and respectively copy the content from the corresponding aggregation. %Otherwise, first observe that we have not yet materialized the skyline tiles and thus we do not have access to the content of each tile. Therefore, we need to find the content of each leaf node from scratch. The simplest way to do this is to run a skyline query for some query point inside each leaf node. However, this is very time consuming. Instead, we use Property \ref{prop:tile_neighbour}. Using this property, we first calculate the skyline for one leaf node, and then, for all the leaf nodes that share a hyper-plane with it, we infer their content based on their neighboring leaf nodes.   

%\subsection{Running Queries}
\textbf{Performing Queries}. At runtime, the query result is determined by a simple traversal of the kd-tree index. The search locates the leaf node that encloses the query, and the list of points stored in that leaf represents the (super)set of the skyline query. 
%Algorithm~\ref{algo:Query} outlines the index traversal process. Note that, 
In case of generalized tiles, the process is run separately for each index structure (i.e., $\lceil n/l \rceil$ times). All searches are completely independent, so the search can be ran in parallel at the SP, thus improving response time.

\subsection{Performance Analysis}

\textbf{Index Construction Time.} Alg.~\ref{algo:getSkylineRegion} first finds the border points of $p$, which takes $O(n^2)$. Then, for each point, it finds the hyper-planes delimiting its skyline region, which takes $O(d^3n^3)$ (line \ref{alg:sky_endpoint} takes $O(dn^2)$ because for any candidate end-point, we need to check if another end-point covers it). Overall, Alg.~\ref{algo:getSkylineRegion} takes $O(d^3n^3)$. Alg.~\ref{algo:getIndex} calls Alg.~\ref{algo:getSkylineRegion} routine for all the points, which costs $O(d^2n^4)$. Then, for each $l$ points, Alg.~\ref{algo:getIndex} builds a kd-tree. Observe that, the height of the kd-tree is $O(d\log N)$, and in total, there are $O(dN)$ hyper-planes. Thus, building the index costs $O(d^2N\log N)$. Then, there are a total of $N^d$ leaf nodes, and filling the content of each takes $O(n)$ which is in total $O(nN^d)$.

\textbf{Query Time.} Searching each index takes $O(d^2\log N)$ (index height is at most $N^d$ and each level requires $O(d)$ comparisons), for a query time of $O(\lceil\frac{n}{l}\rceil d^2\log n)$. 
%On a multi-core server, this can be reduced by a factor linear in the number of cores, since the search is embarrassingly parallel).

\textbf{Space Complexity.}  Every tile can contain $O(l)$ points, and there are $O(\frac{n}{l})$ separate index structures. Therefore, the total space complexity is $O(2^{d^2}n^2l^{d-1})$. 
In general, we can observe that increasing $l$ reduces query time but increases space complexity. Thus, $l$ can be set depending on the space constraints that exist at the service provider.

%Note that the possible locations for the vertical or horizontal lines are only where there already exists a horizontal or vertical line. Each vertical (or horizontal) line corresponds to a segment of the border between a point $p$ and $p'$. Thus, the total number of lines depends, for a point $p$, how many other border points exit. The border points for a given point, $p$, are the skyline points for skyline query at $p$. 
    
%Let $S_{P_i}$ be the skyline points for skyline queries at $P_i$. For a point $P_i$, consider the quantity $\frac{\sum_{P_z\in S_{P_i}}\sum_jI_{P_z\in S_{P_j}}}{|S_{P_i}|}=\frac{\sum_j|S_{P_i}\cap S_{P_j}|}{|S_{P_i}|}$. This quantity measures, on average, for points in $S_{P_i}$, the total number of points in the database for which they are a skyline point. It is, in a sense, a measure of how independent the skyline points $P_i$ are from the skyline points for the rest of the database. Let $c_D=\max_i\frac{\sum_j|S_{P_i}\cap S_{P_j}|}{|S_{P_i}|}$. We expect $c_D$ to be a small constant in practice, although in the worst case it can be equal to $n=|D|$.

%Now consider $N=\sum_i |S_{P_i}|$. For any $j$, we have that $N=\sum_i |S_{P_i}\cap S_{P_j}|+|S_{P_i}\cap (D-S_{P_j})|\leq c_D|S_{P_j}|+sum_i|S_{P_i}\cap (D-S_{P_j})|$ . Choosing another point in $D-S_{P_j})$ and repeating the same process we get that $N=\sum_i |S_{P_i}|\leq c_D\times n$. 
    
%\textit{Average-case bound on N for uniform data distribution}. 

%% file: Encryption.tex
\vspace{-10pt}
\section{Encrypted Skyline Search}\label{sec:encrypteion}
Our result materialization approach reduces the skyline query to a simple index look-up. The benefits of our method become even more clear when performing skyline queries on encrypted data. We do not require any distance calculations at query time, as existing methods do. We only require value comparisons for traversing the index. Furthermore, these comparisons are not performed on the actual data points, but on index node extents. In addition, we bulk-load the indices, which hides any data distribution details, and makes the indexes fully balanced. These features allow us to utilize simple and efficient cryptograpic primitives, while at the same time providing strong security guarantees.

\subsection{Encryption Method}
We need to encrypt a set of skyline indexes $\mathcal{I}$. For each index, we must encrypt (1) the data points stored in the leaf set and (2) the index structure itself. 

\textbf{Encrypting Data Points.} The search does not perform comparisons on data points, so we can use conventional symmetric encryption, such as AES, which provides strong protection and also achieve semnatic security. After traversing the index and reaching the leaf level, we return the entire contents of the leaf to the user, who decrypts them locally.

\textbf{Encrypting Index Structures}. 
%We consider two alternative approaches to protect the index and the values in the internal nodes. 
Since a kd-tree is used, we only need to perform comparisons at each index level. We employ two alternative encryption techniques: mutable order-preserving encryption (mOPE)~\cite{popa2013ideal} and practical order revealing encryption {\em (pORE)}~\cite{chenette2016practical}. mOPE has been proven to be ideal, and does not leak any information about values, (e.g, no value distribution, density, etc.). However, it requires the encoding for each index value to be determined in advance, and the user needs to be aware of this mapping. As a result, the user must perform a one-time setup operation through which it downloads the mapping from the DO. On the other hand, with pORE, the user can compute the ciphertext of an arbitrary data value based on the secret key alone, without any mapping tables. However, pORE incurs a small and measurable leakage in the form of the position of the most significant bit that differs when comparing two values. Recall that, the comparison is not performed directly on data points, but on intermediate index values. Still, this amount of leakage may not be acceptable in some scenarios. Therefore, our two solutions offer a measurable trade-off between protection and setup cost: if the user is willing to download the mapping locally, then the ideal mOPE can be used. Otherwise, if the user is willing to trade a small amount of leakage, then pORE can be used, with no additional one-time setup required.

\subsection{Security Analysis}
We assume the clients do not collude with the server and that the server is honest-but-curious (i.e., it correctly follows the protocol, but tries to infer additional information). %That is, the server always follows the algorithm but tries to learn from about the queries or the data as it follows the algorithm. 
%Since the clients do not collude with the server, the server cannot perform queries on the data where it knows the content of the query. 
Our analysis quantifies the security leakage and answers the following questions: (1) what can the server learn about the data by just observing the skyline index (static data leakage), (2) what can the server learn about the data while performing encrypted queries on the index (dynamic data leakage) and (3) what can the server learn about a query when performing the encrypted query (query security).

\textbf{Static Data Leakage}. The SP observes the index and attempts to learn information from the structure. We need to show that, given a leakage function $\mathcal{L_S}$, the server can only distinguish with negligible probability between a \textit{real} index, $R$, created based on the original data and a \textit{simulated} index, $S$, created based on the leakage function $\mathcal{L}_S$. The security game is to repeatedly show the server a pair $(R_i, S_i)$ for a polynomial number of rounds and let the server guess whether the first index or the second index is the real one (the game is, in essence, similar to that of IND-OCPA \cite{popa2013ideal}). The following theorem quantifies our leakage. 
\vspace{-8pt}
\begin{theorem}
The server can distinguish between a real and simulated index with negligible probability given leakage function $\mathcal{L(\mathcal{I})}=(|\mathcal{I}|, \{\forall_{I\in \mathcal{I}}|\;(h_I, \sigma(I))\})$, where $|\mathcal{I}|$ is the number of indices, $h_I$ is the height of the index $I$ and $\sigma(I)$ is the size of the content of each node in the index $I$. 
\vspace{-8pt}
\end{theorem}

\textit{Proof sketch}. Given the leakage function, and a fixed leaf set size, a simulated index can be constructed to have exactly the same structure. 
%, e.g. the same height and number of internal nodes of the index. 
This results from our proposed index construction method, where we bulk-load the tree, and ensure that all leaves are at the same height. Thus, the security of our method boils down to the security of the underlying order-preserving encryption scheme. % (recall that mOPE provides ideal security).
\qed

The above result is possible because we build a completely balanced tree, and only the order of magnitude for the leaf set is revealed. From an empirical perspective, the height of the index reveals very little about the data, as a wide range of data sizes lead to a skyline index structure of the same height. If leaking the leaf set size is considered unacceptable, one can employ padding, where fake points are added to the leaf nodes. Doing so will not have an effect on search performance, but increases communication cost.

\textbf{Dynamic Data Leakage and Query Security}. We study dynamic data security and query security together as they are both determined by the encryption method used for performing comparisons. 

\textit{mOPE}. If we use mOPE \cite{popa2013ideal}, we can guarantee that the only leakage from an individual query is its traversal path, and that there is no extra leakage (in addition to $\mathcal{L}_S$) from the data. This follows the security analysis of mOPE \cite{popa2013ideal}. Intuitively, this holds because a query is translated into a traversal path of the index and then sent to the server. As a result, the server only learns the path accessed by the query and nothing more. At the same time, the server only accesses the index nodes based on the path provided by the query, which it could have done without the query as well.
% Thus, the leakage about the data is the same as the static case.  

\textit{pORE}. Using pORE \cite{chenette2016practical} increases the leakage of our algorithm but removes the extra storage requirement at the user side because of mOPE. Note that our encryption algorithm sends an independently encrypted value for comparison at each level of the index (that is, the encrypted query may contain the same dimension encrypted multiple times, and the size of the query is equal to the height of the tree). 

Intuitively, this ensures that, at the server side, every comparison is independently done, and the server cannot learn anything by cross-examining the queries. Thus, the leakage is reduced to that of pORE \cite{chenette2016practical}, which is the index of the first bit that is different between the query and the index node. Since the server does not encrypt the query points, this leakage may not have any meaningful implication regarding the security of the data. However, it provides a lower level of guarantee compared with mOPE.

\textit{Protecting Traversal Patterns}. We do not directly protect access patterns to the index. While access patterns themselves may not lead to an adversary learning the data values, they may reveal information about the query. To avoid such disclosure, our approach can be used in conjunction with oblivious RAM structures. There are a number of existing techniques~\cite{oram1, oram2} that can be used in conjunction with generic index structures (including kd-trees) in order to hide access patterns through re-balancing. That will generate additional maintenance cost, although query response time will not be significantly affected. The details of combining ORAM with our approach are not specific to skyline queries, so they fall outside the scope of this submission.
%Based on the leakage discussed above, the access pattern of a query becomes known to the adversary. This is a common problem [XXX] and there is active research in addressing the issue in [XX]. We acknowledge that this can allow adversaries to perform correlation attacks on the clients. For instance, the adversary can know if two different clients received the same results or not, although it cannot know what the result is. 

%% file: Exp.tex
\subsection{Experimental Setup}
We performed experiments on an Intel i9-9980XE CPU (3GHz) with 128GB RAM running Ubuntu 18.04 LTS.

\textbf{Dataset.} Following the setup from~\cite{liu2018secure}, we use both synthetic and real datasets, but with larger sizes. We used the NBA\footnote{Retrieved from https://stats.nba.com/ on on 04/15/2015} dataset with 2438 points in 5 dimensions, where each point represents an NBA player's performance metric (e.g., points scored, blocks, assists, etc.). We use three synthetic datasets: uniform, correlated (Gaussian distribution with correlation coefficient 0.9) and anti-correlated (Gaussian with coefficient -0.9). We consider up to 50,000 points, and dimensionality up to 5.

\textbf{Algorithms.} We evaluate our dynamic programming algorithm (label \texttt{DP}) and greedy algorithm (label \texttt{GREEDY}) from Sec.~\ref{sec:agg}, and for each of them we consider the generalized tiling option discussed in Sec.~\ref{sec:genTiles}. The notation \texttt{GREEDY-l} (respectively \texttt{DP-l}) for some value of \texttt{l} refers to the greedy (respectively DP) algorithm with generalized tiling parameter set to $l$. 
%We also vary the false hits parameter $k$. 
We also include the standalone generalized tiling algorithm (without aggregation, i.e., skyline indexes are built directly on the skyline tiles) with label \texttt{GEN-TILE}.

To the best of our knowledge, the state-of-the-art work for skyline queries on encrypted data is that of \cite{liu2018secure} and~\cite{IoT19}. The former requires two non-colluding servers, and it takes around three hours to complete a query, whereas the latter requires a time in the order of seconds for $100$ data points. Since our method is much faster (sub-second query response time), we did not include a direct comparison with these approaches, which we clearly outperform -- mainly due to our approach of materializing results. %our method with theirs experimentally for the following reasons. Firstly, they assume existence of two non-colluding servers, which is an unrealistic setting and differs with our scenario where only one untrusted server is assumed. Secondly, the query time shown in their experiments is in the order of hours for datasets smaller that the ones used here, which is extremely impractical. In the absence of any existing practical baseline for our scenario, we compare the algorithms mentioned in this paper with various parameter settings. 

\textbf{Measurements.} We report construction time, query time and storage cost. Construction time is the time required to build and encrypt the index structure(s) at the data owner. Query time is the time to execute a query on the encrypted index. Storage cost measures the amount required to store the SP. Result filtering time at the user takes less than half a second, so we omit it from the measurements.

\begin{figure}[t]
\hspace{-0.5cm}
        \centering
        \includegraphics[width=1.05\columnwidth]{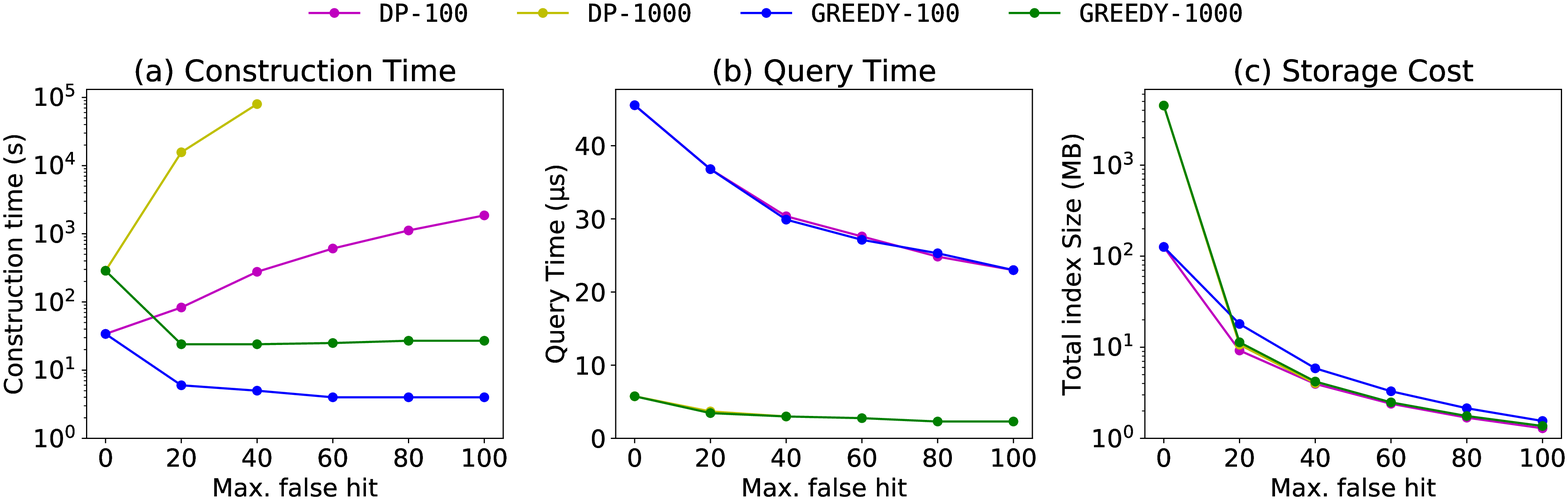}
        \vspace{-25pt}
        \caption{Varying false hits on smaller dataset}
        \vspace{-10pt}
        \label{fig:exp:small_n}
\end{figure}
\begin{figure}[t]
\hspace{-0.5cm}
        \centering
        \includegraphics[width=1.05\columnwidth]{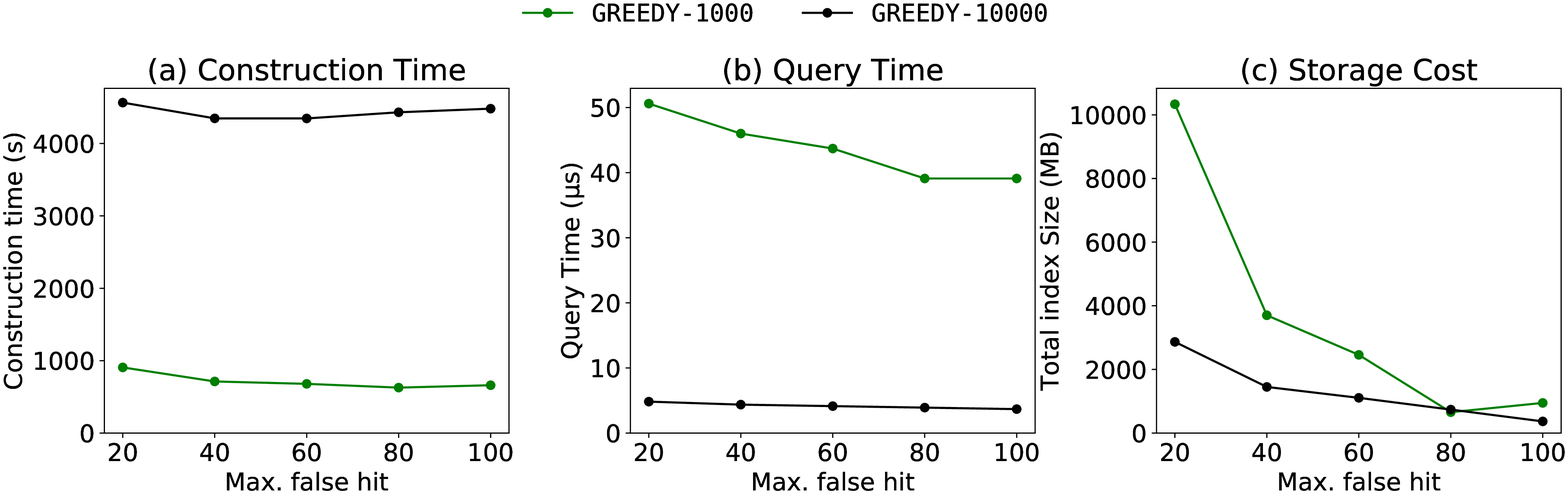}
        \vspace{-20pt}
        \caption{Varying false hit on larger dataset}
        \vspace{-15pt}
        \label{fig:exp:large_n}
\end{figure}
	
\begin{figure}[t]
\hspace{-0.5cm}
        \centering
        \includegraphics[width=1.05\columnwidth]{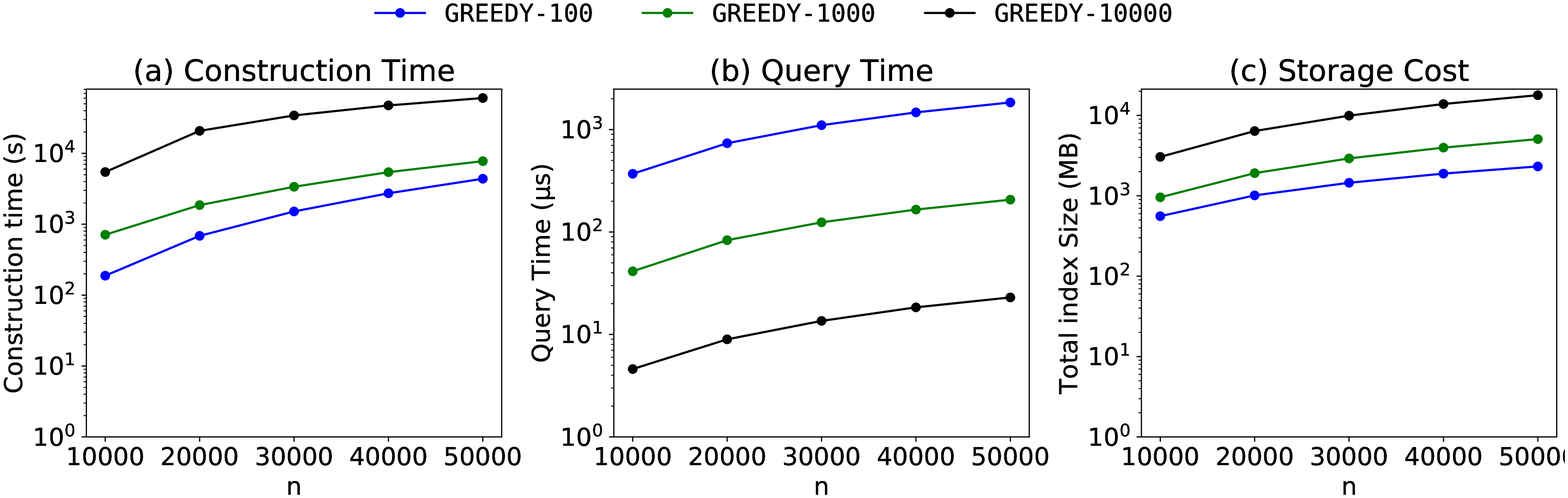}
        \vspace{-25pt}
        \caption{Varying data size}
        \vspace{-10pt}
        \label{fig:exp:change_n}
\end{figure}
\begin{figure}[t]
\hspace{-0.5cm}
        \centering
        \includegraphics[width=1.05\columnwidth]{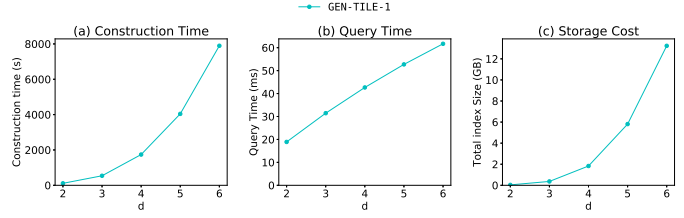}
        \vspace{-20pt}
        \caption{Varying dimensionality}
        \vspace{-20pt}
        \label{fig:exp:change_d}
\end{figure}

\begin{figure}[t]
\hspace{-0.5cm}
        \centering
        \includegraphics[width=1.05\columnwidth]{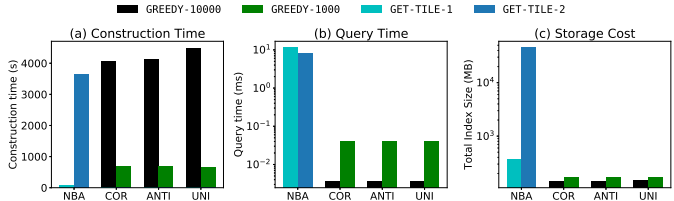}
        \vspace{-25pt}
        \caption{Other Distributions}
        \vspace{-15pt}
        \label{fig:exp:change_dist}
\end{figure}

\subsection{Results on Uniform Data}\label{sec:exp:uniform}
%Here we present our results on datasets with uniform distribution. The majority of our experiments are on uniform data because as will be discussed in Sec. \ref{sec:exp:other_dist}, data distribution does not seem to have a large impact on the performance of our algorithms in practice.

%We present separately results for two-dimensional data and for higher-dimensional data. 

{\bf Experiments on two-dimensional data.}
%\label{sec:exp:2d}
%Due to scalability issued of \texttt{DP}, we use a smaller dataset to perform experiments involving that include this method. \texttt{DP} (in Sec. \ref{sec:exp:small}) and we perform experiments on larger datasets in Sec. \ref{sec:exp:large}.
%\paragraph{Experiments on smaller data}\label{sec:exp:small}
First, we compare the performance of the proposed DP and greedy algorithms for multiple settings of $k$ and $l$. Since the DP approach is costly, we restrict these comparative runs to a dataset of $1,000$ records, and evaluate the greedy approach later on using larger data sizes. Fig.~\ref{fig:exp:small_n} summarizes our findings. Fig~\ref{fig:exp:small_n}(c) shows that although \texttt{GREEDY} may not return optimal solutions, in practice, it returns solutions with storage cost very close to that of \texttt{DP}. However, Fig.~\ref{fig:exp:small_n}(a) shows that \texttt{DP} takes much longer to run due, to its higher time complexity. Fig.~\ref{fig:exp:small_n}(b) shows that the query time is almost the same regardless of whether \texttt{GREEDY} or \texttt{DP} is used. Observe that, the construction time is generally smaller for $l=100$ compared with $l=1000$, while query times are about a multiplicative factor apart ($l=100$ requires $10$ different indexes to be searched while $l=1000$ only searches one index of larger size). We emphasize that the query time, which is the delay perceived by the user at runtime, is very small, always less than one millisecond.

The results show that \texttt{DP} provides a relatively small storage advantage compared with \texttt{GREEDY}, whereas it index construction time overhead is considerably larger. \texttt{DP} becomes impractical for larger values of $k$ even when data set contains only $1,000$ points. In the remainder of the experiments, we exclude \texttt{DP} from our evaluation. However, it remains an interesting approach from a theoretical perspective, and may prove valuable in future work as a base for deriving effective heuristics that reduce storage cost.

%\paragraph{Experiments on larger data}\label{sec:exp:large}
%On larger datasets, we performed two sets of experiments. We studied the impact of data size and number of false hits on the performance of our algorithm. 

%\textbf{Experiment on number of false hits}. 
Next, we increase data size to $n=10,000$ and vary parameter $k$ (which controls the maximum number of false hits allowed). Results are shown in Fig.~\ref{fig:exp:large_n}. In general, $k$ does not impact significantly construction time, and only has a visible impact on query time when $l=1,000$. This is due to the fact that allowing more false hits reduces more significantly the number of tiles created when $l=1,000$, compared with the case when $l=10,000$. We suspect this occurs due to $k$ being smaller (ranges from $2$ to $10$) when $l$ is $1,000$ and increasing it allows for more flexibility of aggregations. Finally, we observed that index structure size is about 50MB for $k=60$, which shows that the communication cost for transferring the structure when using mOPE is small.

%\textbf{Exp on data size} 
We further increase data size up to $50,000$ data points and set $k=\frac{l}{100}$, so that the false hits count is $k\times \frac{n}{l}=\frac{n}{100}$, i.e., $1\%$ of the data. Fig.~\ref{fig:exp:change_n} shows the results. For $l=10,000$ the storage cost and construction time of the index become prohibitive, but the query time is at least an order of magnitude smaller than in other cases. However, smaller values of $l$ can be used in practice to handle this workload.

{\bf Experiments on higher dimensional data.}
%\label{sec:exp:higher_d}
Observing the theoretical results of Sec.~\ref{sec:genTiles}, we only use our generalized tiling approach with no aggregation in this section (i.e., the \texttt{Gen-Tile} algorithm). Our decision is due to the following factors: recall that in Sec.~\ref{sec:genTiles} we discussed decreasing the value of $l$ for higher dimensional data to be able to overcome the curse of dimensionality. At the same time, the number of false hits, whenever $k>0$, increases when decreasing $l$. Moreover, to be able to achieve construction time similar to that for 2D data when dimensionality increases, we need to decrease the value of $l$. As a consequence, we need to set $l$ to a small value, but then setting $k>0$ results in an unacceptably large number of false hits. Therefore, for high dimensional data, we suggest setting $k=0$ and $l$ to a small value. That is, we perform no aggregation, but increase the number of skyline indexes. Thus, in this section, we only report results for \texttt{Gen-Tile}.

Fig.~\ref{fig:exp:change_d} shows the results when $n=10,000$. We set $l=1$ since for $d>4$, larger values of $l$ incur significantly more storage cost. Overall, the query time increases linearly with dimensionality. However, for higher dimensional data, storage cost surpasses $10$GB and makes this approach less applicable for dimensionality higher than five. 

\subsection{Results on Non-Uniform and Real Data}\label{sec:exp:other_dist}
We performed experiments on non-uniform data and real datasets as well. The results are shown in Fig.~\ref{sec:exp:other_dist}. The algorithms performs almost identical when comparing different synthetic distributions. This shows one significant difference between dynamic skyline queries and conventional skyline queries. Since in dynamic skyline queries the query point can be anywhere in the space, an anti-correlated distribution does not necessarily increase the size of the skyline result for a query (whereas in the traditional skyline, number of skyline results may change significantly by changing the data distribution). Finally, on the real NBA dataset, our algorithm can perform skyline queries within milliseconds, and when $l=1$ only requires storage cost of about 300MB.

%% file: Related.tex
%The related work fall into two categories: work on skyline queries on plain text and secure skyline queries.

\textbf{Plain-text Skyline Queries}. The skyline query was first discussed in~\cite{kung1975finding}, and gained significant attention following the more recent work in~\cite{borzsony2001skyline}. Variations of the query under different scenarios have been extensively studied~\cite{dellis2007efficient, papadias2003optimal, sharifzadeh2006spatial, chan2006finding, lian2008monochromatic, liu2015finding, yu2017fast, Kossman02}. The {\em dynamic} skyline query was formalized in \cite{dellis2007efficient}, although general skyline algorithms \cite{papadias2003optimal} were able to answer dynamic skyline queries before that. 

Closer to our work are algorithms focusing on continuous skyline queries for location-based services \cite{lee2009continuous, huang2006continuous, lin2011range, cheema2013safe}. These algorithms find ranges where the answer to the query does not change, and incrementally update the skyline when the answer does change. These algorithms exploit \textit{spatio-temporal coherence}, which focuses on how the query and objects move over time to determine how the result of the skyline query evolves.
% over time and consider the Euclidean distance between objects and the query point as a dynamic attribute. 
This creates a fundamentally different problem, as the dominance relationship in our problem is defined differently, and there is no assumption on how queries change over time. As an example, observe that a data point far from the query (measured by Euclidean distance) will be \textit{spatially dominated}~\cite{sharifzadeh2006spatial} by closer points, but this is not necessarily the case with our dynamic skyline problem. In fact, this lack of \textit{locality} is the main challenge in materializing dynamic skyline queries, as data points far from a query point can impact the result.  

The work in~\cite{liu2018skyline} studied independently from us how to materialize the result of dynamic skyline queries. 
%Lacking our observation that domination regions can be found by the solution to a set of inequalities, 
In~\cite{liu2018skyline}, the space is partitioned into a grid, and grid cells with the same skyline result are merged. This leads to redundant time and space utilization, although the outcome is similar to our skyline tile concept discussed in Sec.~\ref{sec:tiling}. The authors of~\cite{liu2018skyline} proposed various methods for merging the grid cells to obtain the skyline tiles. Our method of finding skyline regions allows us to directly create {\em exact} skyline tiles. More importantly, \cite{liu2018skyline} ignores the large storage cost of full materialization, which is the motivation for our contributions in Secs.~\ref{sec:agg} and~\ref{sec:genTiles}. In contrast to our approach, \cite{liu2018skyline} always performs full materialization, which is highly impractical due to the storage cost. 

\textbf{Secure Skyline Queries}. The powerful trend towards outsourcing data storage and querying~\cite{Haci02} led to a significant body of research on querying encrypted data. Most of this work focused on nearest-neighbor (NN) queries~\cite{Hashem10,ESBJ13, Hu11}, culminating with the work in~\cite{SNN} which showed that the most secure and efficient way to answer NN queries on encrypted data is through materialization of results and encryption of the resulting structure. Our work follows a similar model, but we tackle the dynamic skyline query, for which result materialization is much more challenging.

The problem of securely answering skyline queries only received attention very recently. The work in~\cite{Chen16} was one of the first to address outsourced skyline queries, but it only focused on authentication of results, {\em not} data confidentiality. Skyline queries on encrypted data were first considered in~\cite{FGCS16}, where multiple parties engage in an interactive protocol to execute skyline queries on their joint datasets. Each party has access to its own data in plaintext (e.g., there are multiple data owners, and they all must be online at query time). This setting is considerably less challenging than ours. The work in~\cite{liu2018secure} considers a model with two non-colluding servers that engage in a secure multi-party computation protocol to determine the result of the skyline query. The solution is slow, and the assumption that two non-colluding SP's agree to jointly offer a secure skyline service may not be feasible in practice. Finally, the recent work in~\cite{IoT19} provides a single-SP solution based on fast secure permutations and comparisons. However, the solution relies on bilinear map pairings, which are notoriously expensive. The experimental results in~\cite{IoT19} show performance numbers only up to $200$ data points.

%% file: Conclusion.tex
We proposed the first approach to use query result materialization for answering dynamic skyline queries on encrypted data. Compared to existing work on secure nearest-neighbors, the problem we tackle is much more complex, due to the fact that pre-computing skyline results lacks the locality property that allows NN solutions to be so efficient. We provided an in-depth theoretical analysis of skyline result materialization, and investigated extensively the trade-off that emerges between computational and storage costs. Our proposed heuristics are able to build the result materialization structure in reasonable time, while keeping the storage overhead at practical levels. Our ability to create balanced result materialization structures helps minimize the amount of leakage. In future work, we plan to study additional heuristics that can further reduce construction time. In addition, we will focus on the challenging problem of supporting incremental updates to skyline result materialization, so we can efficiently handle fast-changing datasets.

%% file: Appx.tex
\section{Proofs} \label{sec:appx}

\begin{figure}[t]
    \centering
    \includegraphics[width=\columnwidth]{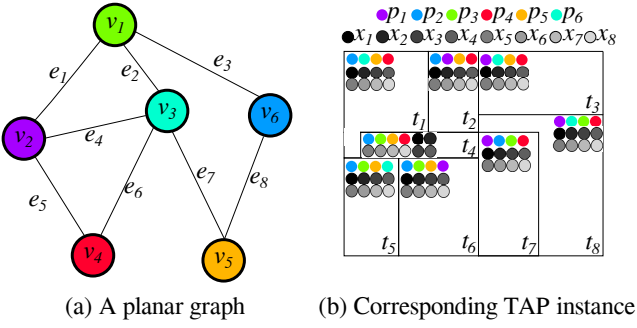}
    \caption{Reduction from PVC to TAP}
    \label{fig:reduction}
\end{figure}

\textit{Proof of Theorem \ref{thm:NPHard}}. We provide a polynomial time reduction from  an instance of planar vertex cover to an instance of TAP. 

\begin{definition}{Planar Vertex Cover (PVC)}
Given a planar graph $G=(E, V)$, find a set of vertices of minimum cordiality, $S\subseteq V$ such that every edge, $e\in E$, there exists a vertex $v\in S$ where $e$ is incident on $v$.
\end{definition}

Recall that and instance of PVC, $\mathcal{I}_{PVC}$ is defined by a graph, $G$, where an instance of TAP, $\mathcal{I}_{TAP}$ is defined by a set of tiles $T$, a database $D$ and an integer $k$ such that the set the content of each tile in $T$ is a subset of $D$. Thus, we describe a polynomial time algorithm that returns $T$ and $k$ given a graph $G$, such that by solving $\mathcal{I}_{TAP}$ optimally we can solve $\mathcal{I}_{PVC}$ optimally as well.

Our reduction works as follows. For each edge in $E$ we create a corresponding tile in $T$ (therefore $|T|=|E|$). For each vertex in $V$ and each edge in $e$ we create a corresponding data point (to be assigned to the tiles, therefore, $|D|=|E|+|V|$).

The reduction has two steps. First, we construct the tiles and then assign points to the tiles. The construction of the tiles is done so that, for any set edges, $B$, incident on a given vertex, the aggregation of the tiles corresponding to $B$ is location-wise valid. We that such a set of tiles can be constructed in polynomial time below.

\begin{lemma}\label{lemma:construction}
Given a planar graph $G=(E, V)$, we can construct, in polynomial time, a set of tiles such that for any set edges, $B$, incident on a given vertex, the aggregation of the tiles corresponding to $B$ is location-wise valid
\end{lemma}

%The above construction ensures that the aggregation of all tiles that correspond to incident edges is location-wise valid. This is because the aggregation can contain edges on the path from the first edge to the second.

Now consider allocating points to the tiles. Consider two sets of points, $D_V=\{p_1, p_2, ..., p_{|V|}\}$, where $p_i$ is a point corresponding to the vertex $v_i$, and the set $D_E=\{x_1, x_2, ..., x_{|E|}\}$, where $x_i$ is a point corresponding to the edge $e_i$. Let $D=D_E\cup D_V$. First, for each tile, we insert the entire set $D_E$ as its content. Furthermore, for a tile $t_i$ corresponding to the edge $e_i=(v_x, v_y)$ we insert the set $D_V\setminus\{p_x, p_y\}$ to its content. Therefore, each tile contains exactly $|D|-2$ points. The intuition behind adding some points of $D_V$ to each tile is to control which tile aggregations are cardinality-wise valid. The intuition behind adding $D_E$ to all the tiles is simply to increase the size of the content of each tile. Doing this enforces TAP to selects the fewest number of aggregations and helps us translate TAP's objective (which is in terms of total number of poitns stored) to VPC's objective (which is in terms of total number of aggregations performed), shown below. Fig. \ref{fig:reduction} shows the reduction for an instance of VPC to an instance of TAP.

Finally set, $k=1$. This enforces that aggregation of two tiles that do not correspond to incident edges is cardinality-wise invalid. This is because any such aggregation will always have exactly $|D|$ points, since it will contain $D_E\cup D_V\setminus\{p_x, p_y\}\cup D_V\setminus\{p_{x'}, p_{y'}\}$. Note that $\{p_{x'}, p_{y'}\}\subseteq D_V\setminus\{p_x, p_y\}$ and therefore, $D_V\setminus\{p_x, p_y\}\cup D_V\setminus\{p_{x'}, p_{y'}\}$ contains both $p_{x'}$ and $p_{y'}$, as well as $D_V\setminus\{p_{x'}, p_{y'}$, which implies it contains the entire $D_V$. Moreover, aggregation of two tiles corresponding to incident edges is cardinality-wise valid. This is because the aggregation contains $D_V\setminus\{p_x, p_{y}\}\cup D_V\setminus\{p_{x}, p_{y'}\}$, which is equal to $D_V\setminus\{p_x\}$. 

Now consider any feasible solution $S$ to the TAP problem. The solution contains aggregations of two types. Firstly, aggregations that contain more than one tile and aggregations that contain exactly one tile. For aggregations corresponding to exactly one tile, $t_i$, consider any one of the two end points of $e_i$, and let $C_1 = \cup\{v_x\}$. For aggregations with more than one tile, all the tiles have to be corresponding to edges incident on a particular vertex, $v_i$. Let $C_2=\cup_x\{v_x\}$. Consider the set $C_S=C_1\cup C_2$. Note that $C$ is a vertex cover of $G$. This is because all tiles are part of some aggregation and the corresponding edges for each tile is covered by the vertex corresponding to the aggregation. In general, for any feasible solution $S$ to TAP we denote the corresponding feasible solution to VPC as $C_S$. Note that we have $|S|=|C_S|$. Therefore, $|S^*|=|C_{S^*}|\geq |C^*|$.

Next, we show that $|S^*|\leq |S_{C^*}|=|C^*|$. Observe that for any feasible solution $C$ to VPC, we can construct a feasible solution $S$ to TAP by just taking, for each vertex in $C$, the aggregation of all tiles corresponding to edges incident on $C$. We denote by $S_C$ the corresponding solution to TAP for a solution $C$ to VPC. Let cost of $S$, denoted by $c(S)$ for a solution $S$ to TAP be the value of the objective function for the solution $S$. First note that since $S^*$ is optimal, $c(S^*)\leq c(S_{C^*})$. Note that $S^*=S_1^*\cup S_2^*$, where $S_1^*$ contains aggregations that contains exactly one tile and $S_2^*$ contains the rest of the aggregations. Then $c(S^*) = |S_1^*|(||D|-2)+|S_2^*|(|D|-1)=(|S_1^*|+|S_2^*|)(||D|-1)-|S_1^*|$ and similarly  $c(S_{C^*})=(|S_1^{C^*}|+|S_2^{C^*}|)(||D|-1)-|S_1^{C^*}|$. Now assume that $|S_{C^*}|<|S^*|$ or $(|S^*|-|S_{C^*}|)\geq 1$. Because $c(S^*)\leq c(S_{C^*})$, we have that

\begin{align*}
(|S_1^*|+|S_2^*|)(|D|-1)-|S_1^*|&\leq (|S_1^{C^*}|+|S_2^{C^*}|)(|D|-1)-|S_1^{C^*}|\\
(|S^*|)(|D|-1)-|S_1^*|&\leq (|S_{C^*}|)(|D|-1)-|S_1^{C^*}|
\end{align*}

Therefore, 
\begin{align*}
|S_1^*|-|S_1^{C^*}|&\geq (|S^*|)(|D|-1)-(|S_{C^*}|)(|D|-1)\\
&\geq (|D|-1)
\end{align*}

Which is a contradiction. Therefore, $|C^*|=|S_{C^*}|\geq|S^*|$. Finally, we get that $|C^*|=|S^*|=|C_{S^*}|$. Therefore, the VPC solution corresponding to $S^*$ is an optimal solution to VPC.

\qed

\begin{figure*}[t]
    \centering
    \includegraphics[width=\textwidth]{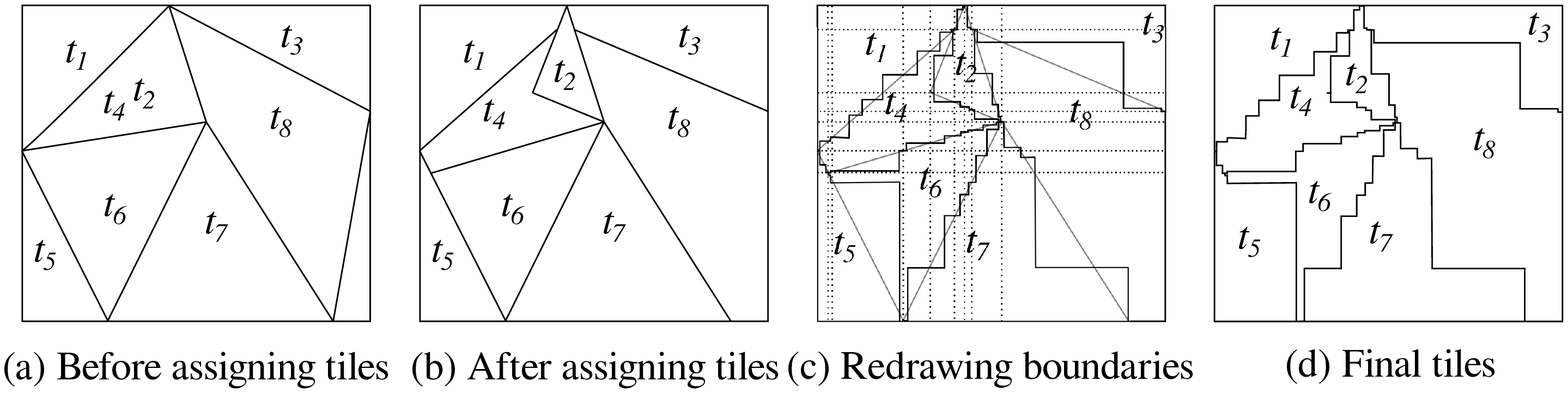}
    \caption{Construction of tiles}
    \label{fig:faces}
\end{figure*}

\textit{Proof of Lemma \ref{lemma:construction}}. The construction works as follows. Construction has two steps. First step is assigning tiles. It works as follows. (1) Draw the planar graph on a plane with all edges as straight lines and draw a minimum bounding rectangle around the graph, (2) Arbitrarily assign each edge to one of the two faces of the graph the edge is a border of, (3) split faces with more than one corresponding edge,  (4) if a face has no assigned edge, remove one of its edges and (5) move the boundaries. Step (1) is straight forward and can be done by Fáry's theorem [X]. Step (2) is also straight forward, and each edge is a border of at most two faces because the graph is planar. 
For step (3), consider  a point inside a face with more than one edge assigned to it, say edges $e_i$ and $e_j$. Connect that point to the both ends of $e_i$ (or $e_j$, the choice of the edge can be done arbitrarily), this creates exactly one new face (because the graph is planar). Assign $e_i$ to this newly created face and remove it from its old assignment. Repeat this process until all faces have at most one point. For example, see $t_2$ in Fig.  \ref{fig:faces}. For step (4) any bordering edge of a face with no points assigned can be removed. The resulting faces will have exactly one point left in them.  For example, see $t_8$ in Fig. \ref{fig:faces}.

The property of this assignment is that for every vertex, there is a corresponding point in space where at least one edge of tiles corresponding to edges the vertex is incident on, ends at that point. This property helps us ensure that incident edges are neighbours. 
The second step is redrawing boundaries. Note that tiles need to have borders that are parallel to the x and y axes, while the current tiles do not satisfy this criterion. To do this, first, consider grid on the space whose vertical and horizontal lines pass through the vertices of the planar graph. Now, in each cell in this grid, a number of lines exist that potentially intersect at the corner of the cells and only at the corner. This is because the graph is planar, and the vertices can only be at the corner of the cells by construction. 

Now in the cell, for each line consider its entrance point and its exit point from the cell. Our goal is to redraw the line such that for each cell, the line's entrance and exit points are the same, but it consists only of segments parallel to x and y axes. To do this, we first shift all the points whose entrance or exit is at a corner arbitrarily away from the corner and add a line segment from the corner to this new point. Now, for all the lines, we start perpendicular to their border and change direction right before, and enter the exit point perpendicular to the location as well. This way, we can ensure no two lines intersect unless they intersect at a corner because they correspond to incident edges. 

Furthermore, observe that borders of all line segments corresponding to edges incident on a vertex intersect at the location of the vertex. Therefore, all the tiles ending at the vertex will be neighbours. However, for the tiles that shouldn't be, we again move their entrance point arbitrarily. Now, there is a path between tiles corresponding to incident edges. 

\qed